\documentclass[12pt]{article}
\usepackage{amsmath,amssymb,amsfonts}
\usepackage[dvips]{graphicx}
\usepackage{epsfig}

\makeatletter \@addtoreset{equation}{section} \makeatother
\makeatletter \@addtoreset{figure}{section} \makeatother

\def\CA{{\cal A}}
\def\CE{{\cal E}}\def\CF{{\cal F}}
\def\CH{{\cal H}}\def\CI{{\cal I}}\def\CJ{{\cal J}}
\def\CK{{\cal K}}\def\CL{{\cal L}}\def\CM{{\cal M}}
\def\CN{{\cal N}}\def\CP{{\cal P}}

\def\CW{{\cal W}}
\def\CZ{{\cal Z}}

\def\g{\gamma}
\def\d{\delta}\def\e{\epsilon}
\def\z{\zeta}

\def\G{\Gamma}

\begin{document}
\begin{titlepage}
\vfill
\begin{flushright}
{\tt\normalsize KIAS-P11033}\\

\end{flushright}
\vfill
\begin{center}
{\Large\bf  Ab Initio Wall-Crossing}

\vskip 1cm

Heeyeon Kim\footnote{\tt hykim@phya.snu.ac.kr}$^\dagger$,
Jaemo Park\footnote{\tt jaemo@postech.ac.kr}$^*$,
Zhaolong Wang\footnote{\tt zlwang@kias.re.kr}$^\diamondsuit$,
and Piljin Yi\footnote{\tt piljin@kias.re.kr}$^\diamondsuit$

\vskip 5mm
$^\dagger${\it Department of Physics and Astronomy, Seoul
National University, \\Seoul 151-147, Korea}
\vskip 3mm $^*${\it Department of Physics \& Center for Theoretical Physics (PCTP),\\
POSTECH, Pohang 790-784, Korea }
\vskip 3mm $^\diamondsuit${\it School of Physics, Korea Institute
for Advanced Study, Seoul 130-722, Korea}

\end{center}
\vfill

\begin{abstract}
\noindent
We derive supersymmetric quantum mechanics of $n$ BPS objects with $3n$
position degrees of freedom and $4n$ fermionic partners with $SO(4)$ R-symmetry.
The potential terms, essential and sufficient for the index problem for
non-threshold BPS states, are universal, and $2(n-1)$ dimensional classical
moduli spaces $\CM_n$ emerge from zero locus of the potential energy. We
emphasize that there is no natural reduction of the quantum mechanics to $\CM_n$,
contrary to the conventional wisdom. Nevertheless, via an index-preserving
deformation that breaks supersymmetry partially, we derive a Dirac
index on $\CM_n$ as the fundamental state counting quantity. This rigorously
fills a missing link in the ``Coulomb phase" wall-crossing formula in literature.
We then impose Bose/Fermi statistics of identical centers, and derive the
general wall-crossing formula, applicable to both BPS black holes
and BPS dyons. Also explained dynamically is how the rational invariant
$\sim\Omega(\beta)/p^2$, appearing repeatedly in wall-crossing formulae,
can be understood as the universal multiplicative factor due to $p$
identical, coincident, yet unbound, BPS particles of charge $\beta$.
Along the way, we also clarify relationships between field theory
state countings and quantum mechanical indices.

\end{abstract}

\vfill
\end{titlepage}

\parskip 0.1 cm
\tableofcontents\newpage
\renewcommand{\thefootnote}{\#\arabic{footnote}}
\setcounter{footnote}{0}

\parskip 0.2 cm

\section{Introduction}

Wall-crossing refers to a phenomenon where BPS
spectrum \cite{Prasad:1975kr,Bogomolny:1975de} changes
abruptly as either a parameter or a vacuum of a supersymmetric field
theory is smoothly changed. In the four dimensional setting, it
was first discovered in the context of $SU(2)$ Seiberg-Witten theory
\cite{Seiberg:1994rs, Seiberg:1994aj},
where the asymptotic spectrum consists of massive vector mesons
and an infinite tower of dyons with unit magnetic charge while the
spectrum at the center of the vacuum moduli space is composed
of a monopole and a dyon \cite{Ferrari:1996sv}.
This apparent ``decay" or ``disappearance"
of BPS states occurs across a wall of marginal stability in the moduli
space,  and this phenomenon turned
out to be generic in all of $N\ge 2$ $D=4$ theories.

The mechanism of the disappearance was clarified a few years later.
It was found that a BPS one-particle state is generically a bound
state consisting of more than one charge centers, which are spatially
distributed according to balance of classical forces
\cite{Lee:1998nv}. Although initial studies were done for 1/4 BPS
states in $N=4$ theories and in classical setting, this was quickly
elevated to a quantum statement \cite{Bak:1999da,Bak:1999ip} and then
extended to $N=2$ theories \cite{Gauntlett:1999vc,Gauntlett:2000ks,Stern:2000ie}. From this new
viewpoint, the wall-crossing occurs simply because the size of
such bound states become infinitely large as a marginal stability wall is approached.
Across the wall, these charge centers become mutually repulsive,
precluding any bound state \cite{Bak:1999da,Gauntlett:1999vc}.\footnote{
These development were, however, restricted to weakly coupled theories,
even though it generalized greatly the previous decades of monopole/dyon
studies in the fully supersymmetric setting of $N=2$ and $N=4$ Yang-Mills
theories. Much of these findings, and their relation to the more conventional
monopole dynamics from 1980's and 1990's, was summarized in a review article
\cite{Weinberg:2006rq}.}
Thus, what was initially thought to be a problem of "decay" is
actually a problem of bound state formation, if viewed backward.

For the simplest $SU(2)$ example, the infinite tower of dyons and the
massive vector meson which exists in the weak coupling side only
should be all realized as loose bound states of the two BPS particles
inside, monopoles and dyons of unit electric charge. At least this must
be the right picture just outside the wall.

The most interesting aspect of this problem is that rules for
the formation or the dissociation of the bound state, whichever way one
view it, seems quite universal and does not depend on detailed dynamics,
which became more clear when the multi-center aspect was rediscovered
later in the $N=2$ black hole context \cite{Denef:2000nb,Denef:2002ru}.
With $N=2$ supersymmetry, BPS states are characterized by the charge
$\gamma$, the central charge $Z_\gamma$, and the supermultiplet.
Thanks to the partially preserved supersymmetry,
the multiplet structure has a reduced form, which can be written as
\begin{equation}
[j]\otimes\left([1/2]\oplus[0]\oplus[0]\right)
\end{equation}
The index that counts degeneracy of such BPS state is the 2nd helicity trace
\begin{equation}
\Omega=-\frac12\,{\rm tr}\left((-1)^{2J_3}(2J_3)^2\right) =
(-1)^{2j}(2j+1) \ .
\end{equation}
For example, when a state $\g_1+\g_2$ disappears across a marginal stability wall,
and dissociates into $\g_1$ and $\g_2$ on the other side, the
indices of these three kind of BPS particles are known to
obey a universal formula \cite{Denef:2007vg},
\begin{equation}
\Omega^-(\g_1+\g_2)=(-1)^{|\langle \g_1,\g_2\rangle|-1}|\langle \g_1,\g_2\rangle|\,
\Omega^+(\g_1)\Omega^+(\g_2) \ ,
\end{equation}
where $\pm$ denote the two sides of the wall.
The wall itself is naturally defined by the condition of vanishing
binding energy that $|Z_1|+|Z_2|=|Z_1+Z_2|$.

This simplest wall-crossing formula has been studied in many
examples, generalized to the so-called semi-primitive cases
for $\g_1+k\g_2$ states \cite{Denef:2007vg}, and most recently embedded into an
algebraic reformulation by Konsevitch and Soibelmann \cite{KS},
which in turn was explained in more physical basis
\cite{Gaiotto:2008cd, Gaiotto:2009hg,  Chen:2010yr, Andriyash:2010qv}.
Despite successes of these,  later and more comprehensive, developments,
much of the literature remain mathematical and sometimes,
from physics viewpoint, opaque. More direct approaches to the
problem, based on physical mechanism of the bound states
and the dissociation thereof, do exist but until very
recently were confined to special examples and situation.

In Ref.~\cite{Lee:2011ph}, a new approach to the low energy dynamics of dyons
in generic $N=2$ Seiberg-Witten theory was proposed. Assuming
that bound states of interest are large, which is always true
whenever the theory is near a  wall of marginal stability, the authors showed how
a $\CN=4$ supersymmetric dynamics can be explicitly written
from the special K\"aehler data of the vacuum moduli space only.
When applied in the limit of a single dynamical probe dyon
in the presence of another (very massive) BPS state, the
bound states can be constructed explicitly and counted,
again confirming the above (so-called primitive) wall-crossing formula.
It is abundantly clear that this method can be used for
an arbitrary number and varieties of dyons, as well, as long
as the proximity to a marginal stability wall is satisfied. In this note, we wish
to set up dynamics of arbitrary number of dyons near such a wall,
with $\CN=4$ supersymmetry,\footnote{To avoid confusion, we will 
use $\CN$ to count quantum mechanical supersymmetries 
and $N$ to count those of  $D=4$ field theory.} and generate wall-crossing formula
via index theorem.

Among the universal formulations that were
previously attempted, Denef's quiver dynamics picture \cite{Denef:2002ru}
gives a very similar picture in the so-called Coulomb phase description.
A recent work by Manschot et.al. \cite{Manschot:2010qz,Manschot:2011xc}\footnote{See
Ref.~\cite{Pioline:2011gf} for a short review.}
fully took advantage of the latter pictures and  wrote down a wall-crossing
formula. At the end of the day, our computation will lead to the same
final wall-crossing formula. Since we start from Seiberg-Witten theory
and derive the wall-crossing formula from scratch, we offer several
improvements.

The first improvement concerns the question of what is the relevant index
theorem. In the Denef's Coulomb phase approach, the most comprehensive studies
to date involve a truncation of dynamics where one ends up with a geometric
quantization problem on the classical moduli space of charge
centers, which are typically compact. In this paper, we denote
such moduli spaces for $n$ centers as $\CM_n$. For two-center case, this
manifold is always $S^2$. The Lagrangian has no kinetic term, but a minimal
coupling to certain magnetic field induces a symplectic
structure on the moduli space, making it a phase space.
In turns out, however, the naive low energy dynamics on this classical
moduli space on $\CM_n$ end up with too many fermionic degrees of freedom.
The anticipated and empirically correct answer, which is a Dirac
index \cite{deBoer:2008zn}, results only if one can somehow remove half of the fermions.
This deficiency has remained unresolved until now.

In this note,
we will explain why the naive truncation to $\CM_n$ was ill-motivated. It turns out
that there is no separation of scales, and all $3n$ bosons and $4n$ fermions
are of equal massgap. Instead, one can choose to reduce the index problem
to $\CM_n$ by deforming the theory with supersymmetry partially broken.
As long as there is one supersymmetry left unbroken and since the
quantum mechanics has a gap, the index is left invariant under the
deformation. At the end of the day, we will thus have provided an ab initio
derivation of the anticipated Dirac index on $\CM_n$, for the first time.

The second concerns the physical interpretation of  certain
rational invariants, defined and extensively used by
Manschot et.al. \cite{Manschot:2010qz}, of the form
\begin{equation}
\bar\Omega(\g)=\sum_{p\vert\g}\frac{\Omega(\g/p)}{p^2}\ ,
\end{equation}
where the sum is over divisors of $\g$. The expression
naturally appears in other formulation of the wall-crossing,
most notably in Konsevitch-Soibelmann. In the course of
enumerating the bound states of bosonic or fermionic statistics,
we will encounter $\Omega(\beta)/p^2$ as a universal effective
degeneracy of $p$ identical particles of charge $\beta$. It
appears as the multiplicative factor from the normal bundle
as one computes contributions from a submanifold fixed by
the permutation group of order $p$.\footnote{This same
numerical factor $1/p^2$ had appeared before in the context of the
D-brane bound state problems of 1990's \cite{Yi:1997eg,Green:1997tn},
where identical nature of the D-branes were also of some importance.}

Along the way, our work also clarifies relation between the field theory
indices, namely the second helicity trace and the protected
spin character, and the quantum mechanical ones. Quantum mechanical
index usually suffers from ambiguity over the definition of $(-1)^F$.
Usual index formulae relies on certain (mathematically) canonical
choice of $(-1)^F$.
Retaining three bosonic coordinates per dyons allow us to inherit
both the spatial rotation group, denoted by $SU(2)_L$, and the
R-symmetry of $N=2$ field theory, $SU(2)_R$. The supersymmetries
belong to $(2,2)$ representation, so both $(-1)^{2J_3}$ of
$SU(2)_L$ and $(-1)^{2I_3}$  of $SU(2)_R$ are chirality operators.
The second helicity trace is then computed unambiguously by
${\rm Tr}(-1)^{2J_3}$. We in turn relate the latter to
${\rm Tr}(-1)^{2I_3}$ which turns out to be equivalent to the
canonical choice leading to the usual Dirac index formula.
This derives, for the first time, the well-known
sign pre-factors in the wall-crossing formulae universally.
In addition, we also explain why the protected
spin character of the field theory is actually computed by
equivariant index, by showing that the quantum mechanical
``angular momentum" operator that appears in the latter is
actually a diagonal sum, $J_3+I_3$, from the spacetime viewpoint.

The paper is organized as follows. Section 2 reviews Ref.~\cite{Lee:2011ph}
and generalize the low energy dynamics to the case of
arbitrary number of dynamical charge centers, and note the
universal nature of the potential terms. Section 3 defines the
index as a method of BPS bound state counting, and in particular
makes contact with the field theory indices, commonly known
as the 2nd helicity trace and its generalization known as the
protected spin character. It turns out that the quantum mechanics found
have $SU(2)_L\times SU(2)_R$ R-symmetry, each of which defines
chirality operators $(-1)^{2J_3}$ and $(-1)^{2I_3}$. The field theory
index corresponds to the former, while mathematical index formulae
are more directly related to the latter. We discuss a universal
relationship between the two, and conjecture that all BPS bound
states in our quantum mechanics are all $SU(2)_R $ singlets.

Section 4 sets up
index theorem for this dynamics and show how reduction to
the classical moduli manifold may be achieved. Here we
show why the naive derivative truncation leading to the
geometric quantization is unjustified by demonstrating that
there is no
natural separation of scales between classically massive
directions and classically massless directions. The main point is that $\CM_n$
is of finite size, and the quantum gaps due to this are
always equal to those along the classically massive directions.
We show, nevertheless, how one can deform  the theory while
preserving the index, such that classically massive modes are
decoupled from the evaluation of the index, at the cost
of partially broken supersymmetry. We also observe that the
reduction process keeps a diagonal subgroup $SU(2)_\CJ$, and
identify the generator $\CJ_3=J_3+I_3$ as the operator usually used
for equivariant index computations. This way, we show that the
equivariant index of quantum mechanics on $\CM_n$ actually
computes the protected spin character of $N=2$ field theory.

After the derivation of Dirac index in section 4, we go on to
evaluate in section 5 the wall-crossing formula by taking into
account the bosonic or the fermionic statistics. Projection operators are introduced
for the purpose,
and the index formula is decomposed into additive
contributions from various fixed submanifolds associated
with  coincident identical particles. The reduced
index problems on the fixed submanifolds appears in the full index
with a universal degeneracy factor $\sim 1/p^2$, which
arises from orbifolding action of the $p$-th order
permutation group $S(p)$. Summing up all relevant contributions,
we find an expression identical to Manschot et.al.'s
wall-crossing formula. Section 6 further supports computation in
section 5, by studying general orbifold index theorem.
We close with summary and comments in section 7.

\section{$\CN=4$ Moduli Mechanics for $n$ BPS Objects}

In Ref.~\cite{Lee:2011ph}, a general framework for deriving moduli
dynamics of dyons of Seiberg-Witten theory was given under
the assumption that one works in the field theory vacuum
where the central charge are almost aligned in terms of
the phases of the respective central charges; in other
words, very near the marginal stability wall.
This program was then carried out explicitly when one
can treat only one dyon as dynamical, with other dyons
as external objects.
In this note, we wish to generalize this to arbitrary number
of charge centers, be they field theory dyons or charged black
holes. For this, all dyons should be
treated as dynamical, and we will denote their charges as
$\g_A$'s.
For the above derivation of one dynamical center,
the proximity to a marginal stability wall played an essential
role, allowing the nonrelativistic approximation and
thus the moduli space approximation possible, so we need to
retain this assumption.

While the moduli dynamics should have $\CN=4$ supersymmetry,
as demanded by the BPS nature of the dyons, simple off-shell
$\CN=4$ descriptions fail to accommodate key interaction terms.
Furthermore, as we will see in section 4 where we compute
the supersymmetric index, it is more convenient to take one
of the four supersymmetries, say $Q_4$, and give up others.
For these reasons, we employ the $\CN=1$ superspace \cite{Maloney:1999dv}
where this supersymmetry, $Q_4$, is manifest. We
package $3n$ bosonic coordinates, $x^{Aa}$, and $4n$ fermionic
superpartners, $\psi^{Aa}$ and $\lambda^A$, as
\begin{equation}
\Phi^{Aa}=x^{Aa}-i\theta\psi^{Aa},\quad
\Lambda^A=i\lambda^A+i\theta b^A \ ,
\end{equation}
with $n$ auxiliary field $b^A$'s. The supertranslation
generator and the supercovariant derivatives are then,
\begin{equation}
Q=\partial_\theta+i\theta\partial_t,\quad D=\partial_\theta
-i\theta\partial_t \ .
\end{equation}

\subsection{Two Centers}

The general structure of two dyon dynamics can be inferred from the results
in Ref.~\cite{Lee:2011ph}. The latter actually derived the effective action of a single
dynamical dyon in the background of an infinitely heavy core BPS state.
When the core state consists of a single dyon, the effective action
derived there can also be regarded as the interacting ``relative" part
$\CL^{rel}$ of a two-dyon effective action, upon the usual decomposition,
\begin{eqnarray}
 {\cal L}= {\cal L}^{c.m.}+{\cal L}^{rel} \ ,
\end{eqnarray}
where the trivial center of mass part was understood to be
\begin{eqnarray}
{\cal L}^{c.m.} =\int  d\theta \;\frac{i}{2}M_{total}
D\Phi^a_{c.m.}\partial_t\Phi^a_{c.m.} -\frac12M_{total}
\Lambda_{c.m.} D\Lambda_{c.m.}\ ,
\end{eqnarray}
with $M_{total}\rightarrow \infty$ understood. Here, let us
recall basic structures of $\CL^{rel}$ as dictated by the supersymmetry.

$\CL^{rel}$ involves only three bosonic coordinates and
four fermionic ones and can be further decomposed as
\begin{eqnarray}\label{super}
 {\cal L}^{rel}&={\cal L}_0^{rel}+{\cal L}_1^{rel}\label{Lrel}\ ,
\end{eqnarray}
where
\begin{eqnarray}
 {\cal L}_0^{rel}&=&\int  d\theta \;\biggl(\frac{i}{2}f(\Phi)
D\Phi^a\partial_t\Phi^a -\frac12f(\Phi)\Lambda D\Lambda
+\frac14\epsilon_{abc}
\partial_af(\Phi)D\Phi^bD\Phi^c\Lambda\biggr) \ ,  \label{L0}
\end{eqnarray}
with $a=1,2,3$, and
\begin{eqnarray}
{\cal L}_1^{rel}=\int d\theta\;\left(i{\cal K}(\Phi)\Lambda -i
{\cal W}(\Phi)_a D\Phi^a\right) \ , \label{L1}
\end{eqnarray}
with the condition  
\begin{equation}\label{n=4condition}
\partial_a{\cal K}=\e_{abc}\,\partial_b{\cal W}_c
\end{equation}
imposed. Note that this also implies
$\partial_{a}\partial_{a}\CK=0$,
which is solved by
\begin{equation}
\CK=\CK(\infty)-\frac{q}{|\vec x|} \ .
\end{equation}
We will see shortly how $\CK(\infty)$ and $q$ can be read off
from the underlying Seiberg-Witten theory.

As was claimed, this Lagrangian is invariant under four
supersymmetries,
\begin{eqnarray} \label{susy-off}
\d_\e x^a&=& i \eta^a_{mn} \e^m \psi^n \ , \qquad
\nonumber\\\d_\e \psi_m &=& \eta^a_{mn} \e^n {\dot x}^a + \e_m b\ ,
\nonumber \\
\d_\e b &=& -i\e_m\dot \psi^m \ ,
\end{eqnarray}
with four Grassman parameters $\e^m$ and with
$\psi^4\equiv\lambda$. The $\CN=1$ superspace we employed is
related to $\e_4$,
so ${\cal L}_0$ and ${\cal L}_1$ are manifestly and individually invariant
under these supersymmetry transformation rules. A less obvious
fact, which is nevertheless true, is that the two are also individually
invariant under all four
\begin{equation}
\d_\e\int dt \, \CL^{rel}_0\; =0\;=\;\d_\e\int dt \,
\CL^{rel}_1,
\end{equation}
if the auxiliary field $b$ is kept off-shell. This is the feature
that allows an easy generalization to $n$ dynamical centers.
The auxiliary field $b$ takes the on-shell value,
\begin{eqnarray}
b=b_{\rm onshell}\equiv\frac{1}{f}\,\left({\cal K} +\frac i4\eta^a_{pq}
\partial_a f \psi^p \psi^q \right) \ ,
\end{eqnarray}
which generates bosonic potential terms of type $\CK^2/2f$ and
mixes up
terms in $\CL_{0,1}$. Nevertheless, $\CN=4$ supersymmetry of
$\CL=\CL_0+\CL_1$ still holds,
now in far more complicated on-shell form.

\subsection{Seiberg-Witten }

Before we extend this to  $n$ dynamical dyons, we need to
understand the role of the core-probe approximation and how
it computes $f$, $\CK$ and $\CW$ \cite{Lee:2011ph}
in terms of the quantities that
appear in the Seiberg-Witten  theory.

Let us consider a collection of charges $\gamma_A$, and
represent it as a semiclassical state.
The basic information about the semiclassical dyon state
comes from the BPS equations of the Seiberg-Witten theory
\cite{Chalmers:1996ya,Mikhailov:1998bx,Ritz:2000xa,Argyres:2001pv}
\begin{eqnarray}\label{BPS}
\vec F_i - i \z^{-1} \vec \nabla \phi_i = 0 \ ,\quad \vec F_{D}^i
  - i \z^{-1} \vec \nabla \phi^i_D =0 \ ,
\end{eqnarray}
where $F=B+iE$ with magnetic field $B$'s and electric field $E$'s,
$\phi$'s are unbroken part of the complex adjoint scalars, each of
which are labeled by the Cartan index $i=1,2,\dots,r$.
$F_D$'s are defined through the low energy $U(1)$ coupling matrix as
\begin{equation}
\vec F_D^i \equiv \tau^{ij} \vec F_j \ , \qquad
\tau^{ji}=\frac{\partial \phi_D^j}{\partial\phi^i}\ .
\end{equation}
The pure phase factor $\zeta$ is determined by the
supersymmetry left unbroken by the charge
$\gamma$ in a given vacuum, and
equals the phase factor of the
central charge $Z_{\g}$ of the configuration.

In a core-probe approximation, we split
$\g_T=\g_h+\sum_{A'}\g_{A'}$
and treat the latter $n-1$ as a fixed background of total charge
$\g_c=\sum_{A'}\g_{A'}$.
As we saw in the previous section,
the Lagrangian for the dynamical dyon (of charge $\g_h$)
is characterized by three objects.

The first is the mass function $f=|\,{\cal Z}_{\g_h}|$ as in
\begin{equation}
\CL=\frac12\, f \left(\frac{d\vec x}{dt}\right)^2 +\cdots\ ,
\end{equation}
where
\begin{equation}
\CZ_{\gamma_h} =\g^e_h\cdot\phi+\g^m_h\cdot\phi_D \ ,
\end{equation}
with the electric part $\g_h^e$ and the magnetic part $\g_h^m$ of
the charge vector $\g_{\gamma_h}$. The scalar fields here solve the above
BPS equation with the other $n-1$ charges $\gamma_{A'}$'s as
the background point-like sources. The fact we treat such dyons
as point-like objects is justified by going very near a marginal
stability wall, since this tends to separate charge centers far
apart from one another. As we will see shortly, this proximity to
marginal stability wall plays a central role in allowing us to
construct nonrelativistic low energy dynamics of dyons.

Clearly $|\CZ_{\gamma_h}|$ acts as the
inertia of the probe dyon, which is position-dependent because
of the background: this sort of identification is in accordance
with general spirit of how one describe well-separated charged
objects \cite{Gibbons:1995yw}, which has been tested and used
successfully for many soliton systems and even lead to exact moduli
space metric in some cases \cite{Lee:1996kz,Weinberg:2006rq}.
We also use the notation
$Z_{\g}$
for the central charge of the charge $\g$ so that $Z_{\g} =
\CZ_{\g}(\infty)$.

The other two, more important for the discussion of BPS bound states,
are the potential ${\cal K}^2/2f$ and the vector potential ${\cal W}$,
so that
\begin{equation}
\CL=\frac12\, f \left(\frac{d\vec x}{dt}\right)^2-\frac{\CK^2}{2f}-
\frac{d\vec x}{dt} \cdot\vec\CW+\cdots\ ,
\end{equation}
where these two are determined entirely by the charge
distribution of $\g_{A'}$'s as \cite{Lee:2011ph}
\begin{equation}
d{\cal W}= *d{\cal K}\ , \quad
{\cal K}={\rm Im}[\zeta^{-1}{\cal Z}_{\g_h}] ={\rm Im}[\zeta^{-1}{ Z}_{\g_h}]
-\sum_{A'} \frac{q_{hA'}}{|\,\vec x-\vec x_{A'}\,|}\ ,
\end{equation}
with\footnote{
This convention for the Schwinger product here follows the one
used by Denef in Ref.~\cite{Denef:2002ru, Denef:2007vg}. The original derivation of dyon dynamics
from Seiberg-Witten theory in Ref.~\cite{Lee:2011ph} used a different convention,
such that
$$\langle\gamma,\gamma'\rangle =\langle \gamma,\gamma'\rangle_{\rm Denef}=2\langle
\tilde\gamma',\tilde\gamma\rangle_{\rm Lee-Yi}.$$
The tilde emphasizes the fact that the latter also used half-integral
electric charges as opposed to integral ones, which is natural when we
compute Coulomb energy. Magnetic charges are integral in either convention.}
$$q_{hA'}=\langle \gamma_h,\gamma_{A'}\rangle/2 \ $$
 for the Schwinger product.

These are
direct consequences of the equations (\ref{BPS}), combined
with the extra assumption of being near the marginal
stability wall. Generically,
the bosonic potential would have been
\begin{equation}
|\CZ_h|-{\rm Re}[\zeta^{-1}\CZ_h]\ ,
\end{equation}
but this reduces to
\begin{equation}
\CK^2/2|\CZ_h|=({\rm Im}[\zeta^{-1}\CZ_h])^2/2|\CZ_h|\ ,
\end{equation}
as we move near
the marginal stability wall defined by alignment of $Z_h$ and $Z_c$ \cite{Lee:2011ph}. The
reason why we need this proximity to the marginal stability wall
is clearly not because of inherent properties of the system, but
rather because of the non-relativistic quantum mechanics approximation
we employed. Far away from the wall, the potential energy would be
not small compared to rest mass of the particles involved, which
will bring dynamics to a relativistic one. However, we do not know
how to handle interacting and relativistic particles at mechanical
level.\footnote{Importance of the wall in the derivation of low
energy dynamics of dyons was also recognized by others \cite{Ritz:2008jf}.}
Nevertheless, this approximation is good enough since we already
know that BPS states are stable far away from marginal stability walls.

An important subtlety we wish to point out here is the choice of $\zeta$.
In the core-probe limit, it appears that $\zeta=\zeta_c=Z_{\g_c}/|Z_{\g_c}|$
is the right choice, since we are treating $\g_h$ as an external
particle in the background given by $\g_c=\sum_{A'}\g_{A'}$.
However, $\zeta$ is tied to the supersymmetry left
unbroken by the configuration and furthermore we are
interested in the supersymmetric bound states of $\g_c$ and
$\g_h$. Around such a state, the low energy dynamics should have
supersymmetries associated with $\g_T=\g_c+\g_h$
rather than those associated with $\g_c$.

One can understand this as capturing the backreaction of the
background due to the probe. Failing to do so clearly will give
us nonsensical answers since, otherwise, the supersymmetry of
the bound state in question would not be aligned with the
supersymmetry of the moduli dynamics.
In the core-probe approximation, the two happen to be the same,
$\zeta_T=Z_{\g_T}/|Z_{\g_T}|=\zeta_c$,  simply because
the total central charge is dominated by that of the infinitely
heavy core state. As we give up the core-probe dichotomy, this
accidental identity will no longer hold, and the preceding discussion
tells us that one must always use $\zeta_T$.

As we give up the core-probe approximation and treat all
charge centers on equal footing, the moduli dynamics will
become quite complicated.
The part of the above action that remains least affected by
this extension is the Lorentz force, coming from $-\dot{\vec x}\cdot \vec\CW$
type couplings. The coefficient $q$ in $\CW$ keeps track
of how one particle's quantized electric (magnetic) charges
see the other particle's quantized magnetic (electric one)
charges. $\CW$ is Dirac-quantized and topological,
and furthermore can arise only from sum of two-body interactions.
Therefore, this part of the interaction can be reliably
computed by adding up all pair-wise Lorentz forces, giving us
\begin{equation}
-\frac{d\vec x}{dt} \cdot\vec \CW \quad\rightarrow \quad
-\frac{d\vec x_A}{dt} \cdot\vec \CW_A
\end{equation}
with
\begin{equation}\label{minimal}
\CW_{Aa} = \sum_{B\neq A} q_{AB}\CW^{Dirac}_a(\vec x_A-\vec x_B) \ ,
\end{equation}
where $q_{AB}=\langle\g_A,\g_B\rangle/2$ and $\CW^{Dirac}$ is
the Wu-Yang vector potential  \cite{Wu:1976ge} of a $4\pi$ flux Dirac monopole.
Note that the $4\pi$ flux of $\CW^{Dirac}$ dovetails nicely with
half-integer-quantized $q_{AB}$, as demanded by the Dirac quantization.

For general $n$ also, $\CN=4$ supersymmetry constrains the Lagrangian
greatly and, as we will see shortly, the potential energy is tied to
such minimal couplings. Knowing the latter will allow us to fix, almost
completely, the analog of $\CK^2/2f$ as well. We will presently
see how this works in $n$ center case. A more difficult
question is how the kinetic terms would generalize, to
which we will only give a general statement rather than precise
solution. In this note, our primary interest is in the
supersymmetric index for non-threshold bound states, which is
independent of details of kinetic term.

\subsection{Many Centers}

For many centers, it is more convenient {\it not} to separate out
the center of mass coordinate.
Let us label the centers by $A=1,2,\dots,n$ and denote their $R^3$
position as $x^{Aa}$ and the charge $\g_A$. The $\CN=1$
superfield content is
\begin{equation}
\Phi^{Aa}=x^{Aa}-i\theta\psi^{Aa}\ ,\quad
\Lambda^A=i\lambda^A+i\theta b^A \ ,
\end{equation}
with $A=1,2,\dots,n$ and $a=1,2,3$.
${\cal N}=4$  transformation rules are,
\begin{eqnarray}
\d_\e x^{A}&=& i \eta^{a}_{mn} \e^m \psi^{An} \ , \qquad
\nonumber\\
\d_\e \psi_m^A &=& \eta^a_{mn} \e^n {\dot x}^{Aa} + \e_m b^A\ ,
\nonumber \\
  \d_\e b^A &=& -i\e_m\dot \psi^{Am}\ ,
\end{eqnarray}
where as before $\psi^{A4}\equiv\lambda^A$.
We again split the Lagrangian into the kinetic part and the
potential part,
\begin{eqnarray}
 {\cal L}&={\cal L}_0+{\cal L}_1\ ,
\end{eqnarray}
and look for $\CL_{0,1}$ separately, with off-shell $b^A$'s.

The $n$-center version of $\CL_1$ is, given (\ref{minimal}),
quite obvious,
\begin{eqnarray}
{\cal L}_1&=& \int d\theta\; \left(i{\cal K}_A(\Phi)\Lambda^A -i{\cal
W}_{Aa}(\Phi) D\Phi^{Aa}\right) \ , \label{L1many}
\end{eqnarray}
since the second term gives precisely the Lorentz force among
dyons and while the first
is induced from the second by $\CN=4$ supersymmetry; One can
check easily that
\begin{equation}
\delta_\e\int dt \,{\cal L}_1=0
\end{equation}
under all four supersymmtries, provided that
\begin{eqnarray}
\partial_{Aa}{\cal
K}_B&=&\frac12\,\epsilon_{abc}\left(\partial_{Ab}{\cal
W}_{Bc}-\partial_{Bc}{\cal W}_{Ab}\right)\ ,
\end{eqnarray}
 and
\begin{eqnarray}\label{n=4condition2}
\epsilon_{abc}\partial_{Ab}\partial_{Bc}\CK_C&=&0\ ,\nonumber\\
\partial_{Aa}\partial_{Ba}\CK_C&=&0\ ,
\end{eqnarray}
for any $A,B,C$. 
We already learned that
$$
{\cal W}_{Aa}=\sum_B \frac{\langle\g_A,\g_B\rangle}{2}{\cal
W}^{Dirac}_a(\vec x^A-\vec x^B)\ ,
$$
 so
$\CK$'s also follow immediately  via the $\CN=4$ constraints as
\begin{equation}
{\cal K}_A={\cal K}_A(\infty)-\frac12\sum_B
\frac{{\langle\g_A,\g_B\rangle}}{|\,\vec x^A-\vec x^B|}\ .
\qquad
\end{equation}
Note that this obeys the constraints except at the submanifold, say
$\Delta\equiv\{x^{Aa}\,:\, \vec x_A=\vec x_B, \, \langle \g_A,\g_B\rangle\neq 0\}$.
The quantum mechanics can be very singular at such places also,
meaning that we should excise $\Delta$
from $R^{3n}$ and impose the regular boundary condition instead.

It remains for us to determine ${\cal K}_A(\infty)$'s.
These  ${\cal K}$'s and $\CW$'s can be traced back to
the original BPS equations (\ref{BPS}), and found by
keeping track of how motion of each center is affected by
the presence of the other $n-1$ centers. After
solving the BPS equations, similarly as in the core-probe limit, we
learn that
\begin{equation}
{\cal K}_A={\rm Im}\left[\zeta^{-1}{\cal Z}_A\right]={\rm
Im}\left[\zeta^{-1}{Z}_A\right]
-\frac12\sum_{B\neq A}\frac{\langle\g_A,\g_B\rangle}{|\,\vec x_A-\vec x_B|}\ ,
\end{equation}
where ${\cal Z}_A$ is computed from the solution to (\ref{BPS}) with
the other $n-1$ charge centers taken as the background but, nevertheless,
with the phase of the total charge,
$\zeta=\sum_AZ_A/|\sum_AZ_A|$, used in the equations.
As we noted above,  this is because we must make sure to use
the supersymmetries that are preserved by the bound state of
all centers. This can be also seen from
${\cal K}_A(\infty)={\rm Im}[\zeta^{-1}Z_A]$, which allows
$\sum_A{\cal K}_A(\infty)=0$ as demanded by the antisymmetric
Schwinger product. Note that
this consistency condition would have been violated if we had used
different $\zeta$'s for different $\CK_A$'s.

The other piece $\CL_0$, containing kinetic terms, is a little
more involved. The simplest way to find the most general $\CL_0$
is via an $\CN=4$ superspace. For this, note that the collection
$\{\Phi^a,\Lambda\}$ can be thought of as dimensional reduction
of a $D=4$ $N=1$ vector superfield
\cite{Ivanov:1990jn, Berezovoj:1991ka}.\footnote{In this version of
$\CN=4$ superspace, $\CL_1$ is not obvious. On the other hand, a
more extended  harmonic superspace form has been found to accommodate
both kinetic terms and potential terms \cite{Ivanov:2003nk}.}
In this map, $x^a$'s come from the spatial part of the vector field,
the fermions from the gaugino, and the auxiliary field $b$ from
that of the $D=4$ $N=1$ vector superfield. See appendix A for more
detail. Here, we are mainly interested in $\CN=1$ form of such
a general $\CL_0$, which is available in Maloney et.al.
\cite{Maloney:1999dv},
\begin{eqnarray}
 {\cal L}_0&=&\int  d\theta \;\frac{i}{2}\,g_{AaBb}
D\Phi^{Aa}\partial_t\Phi^{Bb} -\frac12\, h_{AB}\Lambda^A
D\Lambda^B
-ik_{AaB}\dot\Phi^{Aa}\Lambda^B +\cdots \,
\end{eqnarray}
where the ellipsis denotes four cubic terms that we omit here
for the sake of simplicity. This ${\cal L}_0$ is also invariant
under the four supersymmetries we listed above,
\begin{equation}
\delta_\e\int dt \,{\cal L}_0=0
\end{equation}
on its own with $b^A$'s off-shell, provided that
various coefficient functions derive from a single real function
$L(x)$ of $3n$ variables as
\begin{eqnarray}
g_{AaBb}(\Phi)&=&\left(\delta^e_a\delta_b^f+\e^{\;\;e}_{c\;\;a}\e^{cf}_{\;\;\;
b}\right)
\partial_{Ae}\partial_{Bf}L(\Phi) \ , \nonumber\\
h_{AB}(\Phi)&=&\delta^{ab}\partial_{Aa}\partial_{Bb}L(\Phi)\ ,\nonumber\\
k_{AaB}(\Phi)&=&\e^{ef}_{\;\;\;a}\partial_{Ae}\partial_{Bf}L(\Phi)\ ,
\nonumber\\
&\vdots&
\end{eqnarray}
The $\CN=4$ supersymmetry requires all
those terms as well.
See Appendix A for the complete form of ${\cal L}_0$.

Figuring out the precise form of $L$ for $n$ charge centers
requires further work. For a single dynamical dyon in the
core-probe limit, we know that it is related to the
central charge function as $\partial^2 L=|\,{\cal Z}|$. We expect
that there exists a similarly intuitive generalization for $n$ particles
case as well. In this note,  we are primarily interested in
counting nonthreshold bound state, for which details of $L$ does not
enter.
Determination of $L$ can become an important issue, when we
begin to consider non-primitive charge states. See next subsection
for related comments.

Again, the main point here is that $\CL_0$ and $\CL_1$ are invariant
under the four supersymmetries separately when we keep the
auxiliary fields $b^A$'s off-shell. Combining the two, it follows that
the full Lagrangian
\begin{equation}
{\cal L}_0+{\cal L}_1
\end{equation}
is also invariant under all four supersymmetries.
Integrating out $b^A$'s generates potentials of type $\sim {\cal K}^2$ and mixes
up terms in $\CL_{0}$ and $\CL_1$, but $\CN=4$ supersymmetries of the
entire Lagrangian remain intact.

\subsection{Kinetic Function $L$ : BPS Dyons vs BPS Black Holes}

Note that the potential part  ${\cal L}_1$ of the Lagrangian
looks identical to the similar expression previously found by
Denef \cite{Denef:2002ru}, which has been later used extensively
for counting
BPS black holes bound states \cite{deBoer:2008zn,Manschot:2010qz}.
The latter relied on $\CN=4$ quantum mechanical supersymmetry.
Although we started with Seiberg-Witten theory for the derivation
of $\CL_1$, this part of Lagrangian is entirely determined by
$\CN=4$ supersymmetry combined with long-distance Lorentz forces
among charge centers. Thus appearance of the same $\CL_1$ is
hardly surprising.
In fact, when we apply $\CL_1$ to BPS black holes, it is even
more trustworthy, since the Abelian approximation that would
underlie such an interaction form is valid all the way to horizon.
One cannot say the same for field theory dyons, since at short
distance non-Abelian nature must be taken into account.
Nevertheless, as long as we are near a marginal stability wall
and only long-distance physics matters, it is clear that $\CL_1$
is capable of describing both dyons and black holes.

This does not mean that the moduli dynamics of BPS
dyons and those of BPS black holes are identical.
The difference resides in the kinetic
part $\CL_0$ of the Lagrangian. As demanded by $\CN=4$ supersymmetry,
$\CL_0$ is determined by a single scalar function $L$ of the $n$ position
vectors $\vec x_{A}$.
For instance, $L$  for many BPS
black holes of an identical charge
was found by Maloney et. al. \cite{Maloney:1999dv}
\begin{equation}
L(\vec x_1,\vec x_2,\dots)=-\frac{1}{16\pi}\int dx^3\psi^4
\end{equation}
where $\psi=1+\sum_A (m/|\,\vec x_A -\vec x|)$ with the mass $m$.
On the other hand, for two-center dyon case,  we expect
smooth behavior near $\vec r=0$ \cite{Lee:2011ph} since,
when the mutual distance is small, non-Abelian cores cannot be
ignored and will smooth out Coulombic singularities.
Even if we use the naive Abelian results, $\partial^2L\sim 1/r$
at most. Comparing
this to the two-body case of the supergravity result
shows a substantial difference when the two
objects begin to overlap.

Indeed, there are situations when the two theories are expected
to give different answers.  There is no known example
of $N=2$ field theory dyon which is a bound state of
two or more identical dyons. For black holes, however,
no such restriction seems to exist. If a BPS black hole
of charge $\g$ exist, we expect BPS black holes of charge
$N\g$ also to exist, in fact with large entropy.
In the present context of moduli quantum mechanics,
the latter corresponds to a collection of many charge centers
with many flat directions extending to spatial infinities
and may be realized as threshold bound states thereof.
In such cases, the
kinetic term of the effective action at both short
distances and long distances could be important. This
problem is an important outstanding issue in wall-crossing
phenomena in general, for it provided much-needed input data
on what dyons or black holes are available, to begin with,
to form bound states.

Explicit forms of $L$ for $n$ BPS dyons and for $n$ BPS black holes,
respectively, will be studied in a separate work.

\section{R-Symmetry, Chirality Operators, and Indices }

We wish to compute index of the preceding quantum mechanics
\begin{equation}
{\rm Tr}\left((-1)^Fe^{-sH}\right)\ .
\end{equation}
Since the quantum mechanics is gapped, of which much discussion
will follow in next section, this quantity is truly independent
of the parameter $s$. Thus, following the standard arguments, we
will compute this in small $s$ limit. Before proceeding, however,
it is important to clarify what we mean by the operator $(-1)^F$.
In order for the index to make sense, this operator needs to
anticommute with supercharge(s),
$$\{(-1)^F, Q\}=0\ ,$$
which is the condition needed for
1-1 matching and thus cancelation between bosonic and fermionic states for nonzero
energy eigenvalues. Clearly this is not enough to fix the overall
sign of $(-1)^F$ on the Hilbert space, and an index is also plagued
by this ambiguity. When we compute an index of standard Dirac
operator or de Rham operators, there is usually a canonical choice
that is used widely. We will come back to this, later in next
section, but the choice is a matter of convenience only and, a priori,
has no physical significance.

At field theory level, however, we have an unambiguous and useful
definition of such an index, say, the second helicity trace,
\begin{equation}
\Omega=-\frac12\,{\bf Tr}\left((-1)^{2J_3}(2J_3)^2\right)\ ,
\end{equation}
where the trace ${\bf Tr}$ is over a single particle sector of a given charge.
We wish to fix the sign of the quantum mechanical
index, in accordance with this. Irreducible
BPS multiplets, tensor products of half-hyper-multiplet and
a spin $j$ multiplet, have the index
\begin{equation}
\Omega\left([j]\otimes\left([{ 1/2}] \oplus 2 [{0}]\right)\right) \,
=(-1)^{2j}(2j+1)\ ,
\end{equation}
so often we also write,
\begin{equation}
\Omega={\bf Tr}\left((-1)^{2J_3}\right)\ ,
\end{equation}
with the factored-out half-hypermultiplet understood. This
naturally reduces to the low energy dynamics of
dyons, which then must correspond to an index defined with a
chirality operator that acts exactly like $(-1)^{2J_3}$
\begin{equation}
\Omega\quad\leftrightarrow\quad {\rm Tr}\left((-1)^{2J_3}\right)\ ,
\end{equation}
but of course we need to ask here how such an operator is
realized in the quantum mechanics.

{} As can be inferred from discussions in Ref.~\cite{Lee:2011ph},
the quantum mechanics
of previous section are equipped with $SO(4)=SU(2)_L\times SU(2)_R$
R-symmetry.  This is
easiest to see in how the fermion bilinear couplings to $d\CK$ and $d\CW$
combine to give,
\begin{equation}
 -\frac{i}{2}\,\eta^a_{mn}\,\partial_{Aa}\CK_{B}\,\psi^{Am}\psi^{Bn}
\end{equation}
in the component form, where, as before, $\psi^{Am=1,2,3}=\psi^{Aa=1,2,3}$,
$\psi^{A4}\equiv \lambda^A$, and $\eta$  is the `t Hooft
self-dual symbol. The above form is precise when the metric is
flat, but appropriately  modified preserving $SO(4)$ symmetry when
it is not. For each particle indexed by $A$, bosonic coordinates are
in $(3,1)$ representations while the fermions are in $(2,2)$.
Since  spatial rotations rotate $\vec x^A$ as
3-vectors, $SU(2)_L$ should be interpreted as the rotation group, while
$SU(2)_R$ must be descendant of $SU(2)_R$ R-symmetry of the underlying
Seiberg-Witten theory. The latter rotates only fermions and leaves
the position coordinate intact.\footnote{In the core-probe approximation
of Ref.~\cite{Lee:2011ph}, only
$SU(2)_R$ were generically there, but this was an artefact of treating
some of dyon centers as fixed background.}

In particular, the four supersymmetries are labeled by the $SO(4)$ vector
index, and thus are in $(2,2)$ representations. Denoting generators of
these two $SU(2)$'s by $J$ and $I$, respectively, we thus find
\begin{equation}
\{(-1)^{2J_3}, Q\}\;=\;0\;=\;\{(-1)^{2I_3},Q\} \ .
\end{equation}
The quantum mechanics have two unambiguous and physically meaningful
chirality operators that can be used for index computation. The desired
$(-1)^{2J_3}$ is one of them, therefore, we have an unambiguous way
of computing the field theory index from the low energy quantum mechanics.

On the other hand, there is an interesting and universal relationship
between these pair of chiral operators in the quantum mechanics.
Restricting our attention to the relative part of the low energy
dynamics again, we have
\begin{equation}
(-1)^{2J_3}=(-1)^{\sum_{A<B}\langle\g_A,\g_B\rangle+n-1}(-1)^{2I_3}\ .
\end{equation}
where $n$ is the number of dyons in the low energy dynamics.
This is easy to see by considering how the two $SU(2)$ generators are
constructed in the quantum mechanics. For $SU(2)_R$, which rotate only fermions, we have
\begin{equation}
I_a=\sum_A\left(-\frac{i}{8}\,\epsilon_{abc}\,[\hat\psi^{Ab},\hat\psi^{Ac}]
+\frac{i}{4}\,[\hat\psi^{Aa},\hat\lambda^{A}]\right)\ ,
\end{equation}
where the hat signifies the unit normalized fermion. The spatial rotation
generators
\begin{equation}
J_a=L_a+\sum_A\left(-\frac{i}{8}\,\epsilon_{abc}\,[\hat\psi^{Ab},\hat\psi^{Ac}]
-\frac{i}{4}\,[\hat\psi^{Aa},\hat\lambda^{A}]\right)\ ,
\end{equation}
are similar but differ in two aspects: first, since $SU(2)_L$ rotates $\vec x_A$'s,
the generators include the orbital angular momentum $L$; secondly the fermions
rotate differently, as reflected in the sign of the last term. This latter
difference generates a relative sign between the two chiral operators for
each $(2,2)$ representation of fermions, thus explaining $(-1)^{n-1}$. The
other sign is equally simple, and come from well-known piece of charge-monopole
physics, where the orbital angular momenta is schematically something like
\begin{equation}
\vec L\sim \sum_{A} (\vec x_A \times \vec  \pi_A) +\sum_{A>B}
\frac{\langle\g_A,\g_B\rangle}{2}\frac{\vec x_A-\vec x_B}{|\,\vec x_A-\vec x_B|}
\end{equation}
with the covariantized momenta $\pi_A$.
The orbital angular momentum is constructed from tensor product of spin
$\langle\g_A,\g_B\rangle/2$ representations times usual integral angular
momentum. Then regardless of which particular $SU(2)_L$ multiplet the state
is, integrality vs half-integrality of the orbital angular momentum is
unambiguously determined as
\begin{equation}
(-1)^{2L_3}=(-1)^{\sum_{A>B}\langle\g_A,\g_B\rangle}\ .
\end{equation}
Note that this does not require $\vec L$ being  symmetry operators.

Thus, we have the second helicity trace of $N=2$ dyons which can be
computed via the low energy quantum mechanics as
\begin{equation}\label{sign}
\Omega={\rm Tr}\left((-1)^{2J_3}e^{-sH}\right)=(-1)^{\sum_{A<B}\langle\g_A,\g_B\rangle+n-1}\times
{\rm Tr}\left((-1)^{2I_3}e^{-sH}\right)\ .
\end{equation}
In the subsequent computation, with this relation in mind,
we will eventually identify $(-1)^{2I_3}$ as the canonical
chirality operator $(-1)^{F}$. For this,
there is another sign issue to settle, later when we begin
to quote index formula from literature, since the latter
come with a canonical choice of $(-1)^F$, which may
or may not equal to our choice, $(-1)^{2I_3}$, but we
postpone this to end of next section.

Another reason why $(-1)^{2I_3}$ is useful, even though we ultimately
want $(-1)^{2J_3}$, can be found in the observation \cite{Gaiotto:2010be} that
all explicitly constructed field theory BPS states, to date, are in
$SU(2)_R$ singlets times the universal half-hypermultiplet
(from the center of mass part in quantum mechanics viewpoint).
If this is generally true, we can see that the index with $(-1)^{2I_3}$
is always positive and truly counts the degeneracy.
An interesting question, therefore, is whether in the low energy quantum
mechanics we derived all supersymmetric bound states are $SU(2)_R$ singlets.

An interesting variant of the second helicity trace is the protected
spin character \cite{Gaiotto:2010be},\footnote{We are indebted to Boris Pioline and Jan Manschot
for bringing the question of the protected spin character to our attention.}
\begin{equation}
{\bf Tr}\left( (-1)^{2J_3}y^{2J_3+2I_3}\right)\ ,
\end{equation}
where again we took out the universal half-hypermultiplet from the trace
for simplicity. This clearly reduces to, in quantum mechanics,
 \begin{equation}
{\rm Tr}\left( (-1)^{2J_3}y^{2J_3+2I_3}\right)\ .
\end{equation}
Later we will also see how this quantity is naturally computed,
after we reduce the index problem to the more familiar one that
relies only the classical moduli space $\CK=0$, by the equivariant
index that counts ``angular momentum" representations. As we will
see, this reduction process cannot carry the entire $\CN=4$ supersymmetry,
and, of $SO(4)$ R-symmetry, only a diagonal $SU(2)$ subgroup generated
by $J+I$ survives as global symmetry.
The equivariant index on $\CK=0$ space does not count
representations under spatial rotations but under simultaneous
rotation of spatial $SU(2)_L$ and $N=2$ R-symmetry  $SU(2)_R$.\footnote{Of course,
if the $SU(2)_R$ singlet hypothesis actually holds for the
ground state sector, the end result would not know about $I_3$, anyway.
In fact, on the basis of this hypothesis, this equivalence
was anticipated previously \cite{Manschot:2010qz}. Our
argument in section  4.4 will prove the identity without such an assumption. }
See section 4.4. for more detail.

\section{Index Theorem for Distinguishable Centers }

Now we turn to the problem of counting ground states of
the above quantum mechanics, or equivalently counting BPS
bound states of $n$ dyons. Since the quantum
mechanics has a potential, $\sim \CK^2$, one may expect that
the problem can be reduced naturally to another problem on
the classical moduli space of $2(n-1)$ dimensions, say,
\begin{equation}
{\cal M}_{n} =\{x^{Aa}\;\vert \;{\cal
K}_A=0,\;A=1,2,\dots,n\}/R^3 \ ,
\end{equation}
where the division by $R^3$ is to remove the flat center of
mass part.
This classical moduli space is generically a little more complicated
since some of the centers could be associated with identical
particles, which we will deal with in the next section.

This reduction is not as
straightforward as one might think, however.
Ref.~\cite{deBoer:2008zn}, for example, suggested that one can ignore
the (then unknown) kinetic part of the Lagrangian.
Effectively, in our notation, this would involve a geometric
quantization of $\CL_1$,
\begin{equation}
{\cal L}_{geometric}=\CL_1=-b^A{\cal K}_A- {\cal W}_{Aa} {\dot
x}^{Aa}
+ \frac i2 \partial_{Aa} {\cal K}_B \eta^a_{mn} \psi^{Am}
\psi^{Bn} \ ,
\end{equation}
which is obtained by truncating higher-derivative parts in $\CL_0$.
The auxiliary fields, $b^A$'s, are now Lagrange multipliers,
imposing ${\cal K}_A=0$ as constraints and leaving a lowest
Landau level  problem on ${\cal M}_n$ with the magnetic fields
$\sum_A d{\cal W}_A$.
However, computation of the resulting index, if we take
${\cal L}_{geometric}$ verbatim, generates
wrong results relative to other known spectrum;
The geometric quantization of $\CL_{geometric}$ would
lead to index formula that is known to generate
empirically incorrect answers.

For two body case, for example, the degeneracy $2|q|$ has
been known to be the correct answer for many explicit
constructions. See, for example, Ref.~\cite{Gauntlett:1999vc}
and Ref.~\cite{Lee:2011ph} for explicit two-dyon bound state
construction in the weakly coupled and in the strongly coupled regions of
Seiberg-Witten theory, respectively. On the other hand, the naive lowest
level Landau problem (or equivalently the geometric quantization
problem) gives $2|q|+1$. One would hope that the effect
of fermions in ${\cal L}_{geometric}$ will fix this, but this
apparently does not happen.

The truncation of the kinetic terms in the presence
of fermions is quite subtle, since while bosons acquire
a symplectic structure thanks to the magnetic field, there
is no such analog for fermions. Setting the kinetic term
of fermions to zero will cause the canonical commutator ill-defined,
making the whole reduction process  ambiguous.
One can try to reinstate kinetic terms on $\CM_n$ as a
regulator, but then, the main issue is that the number of
fermions in ${\cal L}_1$ is $4(n-1)$ real while the number
of bosons is $2(n-1)$ real, and these lead to de Rham cohomology
problem on ${\cal M}_n$. For not too small $q$ and when $\CM_n$
is K\"aehler, for example, the index of such a quantum mechanics
coincides precisely with the  state counting of the
bosonic geometric quantization problem,\footnote{ See for example Ref.~\cite{Douglas:2008pz},
where in effect a regularized version of these problems were considered
with kinetic terms on $\CM_n$ and for its fermionic partners present.} again
giving us wrong result for the index.

Really at the heart of the problem is, however, the fact that the
classical massive directions are in fact no more massive
than the classically massless directions. Because the classical
moduli space $\CM_n$ is of finite size\footnote{There are also
some exotic cases corresponding to the scaling solutions.
In these cases, the moduli space is non-compact, from short
distance side, but its volume
in the naive flat metric is still finite.},
it comes with various gaps at quantum level, and it so
happens that these quantum gaps are one-to-one matched and identical
to the gaps associated with the classically massive directions:
the dynamics cannot be really split into two distinct sectors
of heavy and light modes, at all, and contrary to initial
expectation, the reduction to $\CM_n$ cannot be justified.
In fact, this lack of separation of scales is easiest to see
in how fermions enter the Hamiltonian. Half of fermions get mass
from $d\CK$ while the other hand get mass from $d\CW$. However, $\CN=4$
supersymmetry of the quantum mechanics
tells us that the two are one and the same object,
and fermions coupling to $d\CK$ are no more heavier than those coupling to $d\CW$.

Fortunately, we can still decouple these classically massive
directions in the computation of the index problem. This involves
a deformation that breaks all but one supersymmetry, yet because
the quantum mechanics is gapped and the surviving supercharge is
effectively a Fredholm operator, it can be done while preserving
the index. Later in the
section and in Appendix B, we  explicitly show that, as far as
computation of the index goes,
we may reduce the moduli dynamics to an effective $\CN=1$
supersymmetric quantum mechanics with target ${\cal M}_n$,
\begin{equation}
{\cal L}_{\rm for \; index\; only}^{\CN=1}({\CM_n})=\frac12\,
G_{\mu\nu}\dot z^\mu\dot z^\nu
+\frac{i}2 \,G_{\mu\nu} \psi^\mu\dot \psi^\nu+\cdots-{\cal
A}_\mu\dot z^\mu+\cdots\ ,
\end{equation}
where $\CA$ is a gauge field on $\CM_n$ such that
\begin{equation}
d\CA={\cal F}\equiv d\left(\sum_A
\CW_{Aa}dx^{Aa}\right)\Biggl\vert_{\CM_n}\ .
\end{equation}
and $G$ is the induced metric on $\CM_n$.
This Lagrangian must be used only for the purpose of
computing index.

The key point here is that the number of fermions is exactly
half of that in ${\cal L}_{geometric}$. Since
these fermions live on the tangent bundle of $\CM_n$, we have
a nonlinear sigma model with real fermions.
The relevant wavefunctions are spinors on ${\cal M}_n$ and the index
in question becomes a Dirac index,
\begin{equation}
\CI_n(\{\g_A\})=
\int_{{\cal M}_n}Ch({\CF})\hat{A}({\cal M}_n)=
\int_{{\cal M}_n}Ch({\CF})
\end{equation}
with the Chern character $Ch$ of $\CF$. $\hat A$ is
the A-roof genus of the tangent bundle, which will be shown
to be trivial for all $\CM_n$'s.
This formula counts the index when we view individual charge
centers as distinguishable; in section 5, we will extend the
formula appropriately when identical particles are involved
and along the way see why the rational invariants of the
form $\sim \Omega/p^2$ with integer $p>1$  appears in various
wall-crossing formulae.

The Dirac index found here is consistent with de Boer et. al.'s
observation \cite{deBoer:2008zn}
that empirically correct answers  emerge for $n=2$ and $n=3$ if
one assumes that the relevant quantum mechanics admit
spinors on ${\cal M}_n$ as the wavefunction.
This can be then generalized to the refined index
(or equivariant index) and make contact with a series
of recent works by Manschot et.al \cite{Manschot:2010qz,Manschot:2011xc}.

\subsection{Two Centers: Reduction to $S^2$}

Supersymmetric ground states were found and counted
for $n=2$ case in Ref.~\cite{Lee:2011ph}, which gave the correct answer
of $2q$ at the end of the day.  As expected, the wavefunctions are all
maximized near the classical ``true" moduli space ${\cal K}=0$, which
was nothing but a two-sphere threaded by a flux of $4\pi q$. However,
the wavefunctions can also be seen to be very diffuse, too much so to let
us call it ``localized" there.

Here, we will illustrate why a naive reduction to ${\cal M}_2=S^2$
by throwing away entire kinetic term is wrong. After the
latter procedure, one ends up with $\CL_{geometric}$ for which we
need to either geometrically quantize over $S^2$ or
regularize the dynamics by reinserting kinetic term on $S^2$
and concentrate on the lowest Landau level. If we follow
the second viewpoint, we end up with a two dimensional
nonlinear sigma model with four real fermions, so effectively
we will have thrown away only the bosonic radial  coordinate
from the original moduli dynamics.

Let us consider the zero point energy of the relative part of the
two-center mechanics. Three bosons can be split into ``radial"
directions, on which ${\cal K}$ and the mass function $f$
depend, and flat ``angular" directions. With ${\cal K}=a-q/r$
and positive $a$ and $q$, the ground
state is at $r=r_0=q/a$, and the radial direction becomes
a harmonic oscillator of frequency $w={a^2}/{f(r_0)q}$, so
\begin{equation}
E_{radial}\simeq \left(m^{radial}_b+\frac12\right)
w\ge\frac{a^2}{2f(r_0)q} \ .
\end{equation}
The angular part, although classical flat, also comes
with a gap due to the finite volume, and the energy
quantization there goes as
\begin{equation}
E_{sphere}\simeq \frac{\vec L^2 -q^2}{2f(r_0)r_0^2} \ge
\frac{a^2}{2f(r_0)q} \ ,
\end{equation}
since the angular momentum is bounded
below, in the presence of the flux, by $q$. The four real
fermions are paired up into two fermionic oscillators of the
same frequency $w$ as above, so we get contribution from
the fermion sector as
\begin{equation}
E_{fermion}\simeq \left(m_{f}+m_{f}'-1\right) w\ge
-\frac{a^2}{f(r_0)q} \ ,
\end{equation}
where we again see that there is only one scale in the fermion
sector also. Of course, the behavior of fermionic degrees of
freedom must be the same as the bosonic ones, since we have
supersymmetry.

This shows that, without further deformation, the gap of
the classically massive radial direction is exactly the
same as the rest of the degrees of freedom.
If we wish to localize the problem to ${\cal M}_2=S^2$ by
removing the radial mode, we must do something else
so that the gap along the radial direction and the gap
along ${\cal M}_2$ are different, but this seems
impossible under the $\CN=4$ supersymmetry of the quantum mechanics.

Let us remember here that, for the evaluation of index,
one needs only two things: a Dirac operator of some kind
and a chirality operator that anticommutes with it.
One would like to compute the index
\begin{equation}
{\rm Tr}(-1)^Fe^{-s H}\ ,
\end{equation}
for interacting part of the theory. Let us, for the sake
of definiteness, take $H=Q_4^2$, and evaluate
\begin{equation}
{\rm Tr}(-1)^Fe^{-s Q_4^2}\ .
\end{equation}
$\CN=4$ supersymmetry is useful since it constrains
dynamics but all of them are not really necessary to define
an index. It is clear that, as long as we preserve this quantity,
we can even break $\CN=4$ supersymmetry.

Of course, ${\cal N}=4$ supersymmetry is important
when it comes to generating correct supermultiplet
structure to the bound state,  but that only concerns
the free center of mass part. The index must be
computed from relative interacting part of the dynamics,
only for which we will break $\CN=4$ supersymmetry.

Thus we are motivated to give up $d{\cal K}=*d{\cal W}$
condition, thereby keeping only $Q=Q_4$ unbroken.
Let us replace
\begin{equation}
\CK\rightarrow\xi\CK
\end{equation}
with some arbitrarily large
number $\xi$ while keeping $\CW$ as it is.
The ground state energy counting is now
\begin{equation}
E_{radial}+E_{sphere}+E_{fermion}\ge \frac{\xi
w}{2}+\frac{w}{2}-\frac{\xi w+w}{2} \ ,
\end{equation}
since the half of the fermions ($\lambda$ and $\psi^r$) get the
mass from $d(\xi{\cal K})$ and the other half from $d{\cal W}$.
The angular momentum sector mass-gap, $w/2=q/2f(r_0)r_0^2$,
is unchanged since the classical vacuum, $\CK=0$ and
thus the radius $r_0$, and $\CW$ are intact under this rescaling.

It is not difficult to see that the reduced dynamics,
after integrating out heavy modes, is a $\CN=1$ nonlinear sigma
model onto $\CM_2=S^2$ coupled to an external vector field $\CW$.
See Appendix B for complete detail of the reduction process.
We note that since $\CM_2=S^2$  happens to be K\"aehler, the
unbroken supersymmetry gets accidentally extended to $\CN=2$,
although this is not  important for our purpose.

\subsection{Many Centers:  Reduction to ${\cal M}_n$ }

Similarly, we wish to deform the theory by rescaling
$\CK_A\rightarrow \xi\CK_A$, when we have many
dynamical charge centers, as well
\begin{eqnarray}\label{deformed'}
{\cal L}_{\rm deformed}'&=&\int d\theta
\;\biggl(\frac{i}{2}\,g_{AaBb}
D\Phi^{Aa}\partial_t\Phi^{Bb} -\frac12\, h_{AB}\Lambda^A
D\Lambda^B-ik_{AaB}\dot\Phi^{Aa}\Lambda^B +\cdots \nonumber\\
&&\hskip 1cm +i\xi{\cal K}_A(\Phi)\Lambda^A -i {\cal
W}_{Aa}(\Phi) D\Phi^{Aa}\biggr) \ ,
\end{eqnarray}
where $\xi$ is an arbitrarily large number.
As in the two-center case, the bosonic potentials are quadratic
in $\CK_A$'s and there are $n-1$ ``radial" directions that
are of  mass $\sim\xi$. There are also $2(n-1)$ fermions
that couple to $d(\xi\CK_A)$'s, so they are also of mass $\sim\xi$.
The two sets can be decoupled together, thereby reducing
the index problem to ${\cal M}_n$ with real fermions.
It leaves behind a $\CN=1$ supersymmetric quantum mechanics onto
${\cal M}_n$ with $2(n-1)$ bosons and $2(n-1)$ real fermions.
The process does not affect the free center of mass part, so the
latter still comes with 3 bosonic coordinates and 4 fermionic ones.

We may further deform the kinetic part, ${\cal L}_0$,
by taking the simplest form of the kinetic function,
\begin{equation}
L= \frac{1}{2}\,\sum_{A}m_A \vec x^{A}\cdot \vec x^{A}\ ,
\end{equation}
which amounts to
\begin{equation}
g_{AaBb}=\delta_{AB}\delta_{ab}m_A,\qquad
h_{AB}=\delta_{AB}m_A,\qquad k_{AaB}=0,
\end{equation}
and setting cubic terms to zero as well. The simplest
way to justify this deformation is that the kinetic function
approaches this flat metric when distances between charge
centers approach infinity. This asymptotic form is more
than good enough since we can always tune the field theory
vacuum, so that we stay arbitrarily near the marginal stability
wall. There, ${\rm Im}[\zeta^{-1}Z_A]$ approaches zero, and
the submanifold $\CM_n$ is arbitrarily large. Since the index
cannot change under the continuous and sign-preserving
deformation
of ${\rm Im}[\zeta^{-1}Z_A]$, and since the
ambient metric is effectively flat for large $\CM_n$,
the index will be unaffected by this choice of metric.

This leaves us with a very simple $\CN=1$ quantum mechanics
\begin{eqnarray}\label{deformed}
{\cal L}_{\rm deformed}&=&\int
d\theta\;\biggl(\frac{i}{2}\,m_AD\Phi^{Aa}\partial_t\Phi^{Aa}
-\frac12\, m_A\Lambda^A
D\Lambda^A
\nonumber\\
&&\hskip 1cm+i\xi{\cal K}_A(\Phi)\Lambda^A -i {\cal
W}_{Aa}(\Phi) D\Phi^{Aa}\biggr) \ ,
\end{eqnarray}
with target $R^{3n}$ modulo submanifolds given by $\CK_A=\pm
\infty$.
Of this, the free center of mass positions $R^3$ and the accompanying
four real fermions decouples, leaving behind the interacting
part of the moduli dynamics onto $R^{3(n-1)}$. This free part is
also essential since it generates the basic BPS multiplet structure
(whose content equals half of a hypermultiplet) to the bound state.
Then, by taking $\xi \rightarrow\infty$, we decouple $n-1$
``radial" directions
and $2(n-1)$ accompanying heavy fermions, and end up with a
nonlinear sigma model onto $\CM_n$ with real $2(n-1)$ fermionic
partners. See appendix for detailed derivation of this fact.

Thus, we  arrive at the effective Lagrangian, which can be used for
the purpose of computing the index of the original $n$ center problem,
\begin{equation}
{\cal L}_{\rm for \; index\; only}^{\CN=1}= \frac12\,
G_{\mu\nu}\dot z^\mu\dot z^\nu
-{\cal A}_\mu\dot z^\mu
+\frac{i}2\, G_{\mu\nu} \psi^\mu\dot \psi^\nu
+\frac{i}2 \,G_{\mu\nu} \psi^\mu\dot
z^\lambda\G^\nu_{\lambda\beta} \psi^\beta
+\frac{i}{2}\,\CF_{\mu\nu}\psi^\mu\psi^\nu
\end{equation}
again with the induced metric $G$ on $\CM_n$ and, as we already noted,
\begin{equation}
d\CA={\cal F}\equiv d\left(\sum_A
\CW_{Aa}dx^{Aa}\right)\Biggl\vert_{\CM_n}\ .
\end{equation}
Since each $\CW_A$ is a sum of Dirac monopoles at $\vec x=\vec
x_B$'s, we find
\begin{eqnarray}
{\cal F}&=& d\left(\sum_A \sum_{B\neq
A}\frac{\langle\g_A,\g_B\rangle}{2}\,
\CW_{a}^{Dirac}(\vec x_A-\vec
x_B)\,dx^{Aa}\right)\Biggl\vert_{\CM_n}
\cr\cr
&=&d \left(\sum_{A>B} \frac{\langle\g_A,\g_B\rangle}{2}\,
\CW_{a}^{Dirac}(\vec x_A-\vec
x_B)\,d(x^{Aa}-x^{Ba})\right)\Biggl\vert_{\CM_n}
\cr\cr
&=&\sum_{A>B}\frac{\langle\g_A,\g_B\rangle}{2}\,{\cal
F}^{Dirac}({\vec x_A-\vec x_B})\ ,
\end{eqnarray}
where $\CF^{Dirac}$ is the Dirac monopole of flux $4\pi$.
Of four supercharges, $Q_4$ survives the deformation process
above, which is then further reduced to $Q_{\CM_n}$ as heavy
modes are integrated out.

\subsection{Index  for $n$ Distinguishable Centers}

Since this is the plain old nonlinear sigma model twisted by
the minimal coupling to $\CA$, the reduced supercharge is
represented geometrically as the Dirac operator with a $U(1)$
gauge field
\begin{equation}
Q_4\quad\rightarrow\quad Q_{\CM_n}=\gamma^\mu \left(
i\nabla_\mu+\CA_\mu\right) \ ,
\end{equation}
whose index, according to Atiyah-Singer index theorem, is given by
\begin{equation}
\CI_n(\{\g_A\})={\rm Tr}\left((-1)^{F_{\CM_n}}e^{-s Q^2}\right)
=\int_{{\cal M}_n}Ch({\cal F})\hat A({\cal M}_n)\ ,
\nonumber
\end{equation}
as promised, where we must assume a canonical choice of the
chirality operator. This is,
$$(-1)^{F_{\CM_n}}=(2i)^{n-1}\hat\psi^1\cdots\hat\psi^{2(n-1)}\ ,$$
in terms of properly normalized and ordered fermions. See next
subsection for how this choice squares off with physically motivated
chirality operators $(-1)^{2J_3}$ and $(-1)^{2I_3}$ of section 3
and how the latter chirality operators reduce on $\CM_n$.

Curiously enough, the A-roof genus $\hat A$ does not contribute
to the index, thanks to the simple topology of $\CM_n$.
To see this, let us first note that the ambient space,
in which ${\cal M}_n$ is embedded is essentially $R^{3n}$.
For instance, take $\vec x_1=0$ to remove the translation invariance
and make the ambient space $R^{3(n-1)}$, and then
impose $\CK_A=0$, of which $n-1$ are linearly independent.
Therefore, $\CM_n$ is a complete intersection in $R^{3(n-1)}$.
Since A-roof genus is a multiplicative class, we have an identity,
\begin{equation}
\hat A(T\CM_n)\hat A(N\CM_n)=\hat
A\left(TR^{3(n-1)}\bigl\vert_{\CM_n}\right)=1
\end{equation}
among the tangent and the normal bundles. However,
$d\CK_A$'s are nowhere vanishing normal vectors on $\CM_n$,
and thus the normal bundle $N\CM_n$ is also topologically
trivial\footnote{We are indebted to Bumsig Kim for pointing
this out to us.}, and
\begin{equation}
 \hat A(T\CM_n)=1\ .
\end{equation}
It is important to note that this decoupling depends only on
the topology of the ambient space, namely the original $3n$
dimensional moduli space, near the surface $\CK_A=0$.\footnote{
Note that similar argument will not lead to triviality of
other multiplicative class since typically they require
complex bundles in order to be defined. For instance,
$Td(\CM_{2k})$ or $c(\CM_{2k})$ cannot be argued to be
trivial in this manner, for instance, since the normal bundle of $\CM_{2k}$
inside the relative position space $R^{3(2k-1)}$ is of odd
dimension and, if irreducible, cannot be complex.
In particular $\CM_2=S^2$ has a real line as the fibre
when embedded into $R^3$, which is consistent with the nontrivial $c_1$.}
This triviality is also implicit in the explicit evaluation
of these Dirac indices
in Ref.~\cite{Manschot:2011xc}. Although  it turned 
out there that $\hat A(T\CM_n)$ factor did make an important 
difference for evaluation of the equivariant index,
the non-equivariant limit is consistent with trivial $\hat A$.

\subsection{Reduced Symmetry, Index, and Internal Degeneracy}

Since we arrived at the nonlinear sigma-model on $\CM_n$ only
after the deformation of the dynamics, which in particular
removes the extended supersymmetry, we must first ask whether
various operators survive this procedure of deformation and
the subsequent reduction process $\xi\rightarrow\infty$. Of the four original
supersymmetries, $Q_4$ survives the deformation. It's on-shell
form will be smoothly deformed as well, which goes like
\begin{equation}
Q_4=\cdots+\lambda^A\CK^A\quad\rightarrow\quad
Q_4=\cdots+\xi\lambda^A\CK^A \ .
\end{equation}
The ellipsis denotes parts
unaffected by the deformation. We emphasize again that this supersymmetry
is explicitly preserved since the deformed Lagrangian (\ref{deformed'})
is written in the superspace associated with $Q_4$. Then, given
that $Q_4$ is a gapped elliptic operator, at $\xi=1$, this deformation
preserves the index as we increase $\xi$ \cite{Stern:2000ie}.
This $Q_4$ reduces to $Q_{\CM_n}$
of the nonlinear sigma model on $\CM_n$, and obviously the
Hamiltonian, $Q_4^2/2$, gets similarly deformed and eventually
reduced to the natural one on $\CM_n$.

This leaves the global symmetry operators and the chirality
operators. With the $\CN=4$ supersymmetry partially broken, the
$SO(4)$ R-symmetry can be easily seen to be broken.  On the other
hand, the deformation commutes with rotation of $\vec x_A$'s, so
we expect to see some $SU(2)$ symmetry does survive the process.
The question is which  $SU(2)$ in $SO(4)=SU(2)_L\times SU(2)_R$
remains unbroken. The answer is the diagonal subgroup, $SU(2)_\CJ$,
generated by
\begin{equation}
\CJ_a=J_a+I_a\ .
\end{equation}
One can see this in several different ways.

Firstly,  both
$J$ $(SU(2)_L)$ and $I$ $(SU(2)_R)$ are broken by themselves, since they both act
nontrivially on heavy fermion sector.
The diagonal generators $\CJ$'s, on the other hand does not involve $\lambda$
fermions and leave the heavy sector ground state untouched. Secondly, after
deformation and reduction to $\CM_n$, the dynamics is a nonlinear
sigma model, where fermions transform identically to bosons. Recall
that bosons and fermions used to belong to $(3,1)$ and $(2,2)$ of
$SU(2)_L\times SU(2)_R$. In the reduced dynamics, symmetry properties
of the bosons and fermions cannot be different, and indeed under
the diagonal subgroup, bosons and fermions transform identically.
Finally, after the deformation, the dynamics has only one real
supersymmetry $Q_4$ so no R-symmetry is expected. However this
supercharge originates from a $(2,2)$ multiplet under $SU(2)_L\times SU(2)_R$,
so has to transform nontrivially under either of the two individually.
On the other hand, becaues $\CJ$ does not rotate $\lambda$'s, $\CJ$
commutes with $Q_4$ and also with its reduced version $Q_{\CM_n}$. At the level
of reduced dynamics on $\CM_n$, this $SU(2)_{\CJ}$ is not an R-symmetry
but a global symmetry that arises from the universal isometry of $\CM_n$.

While we are on the question of symmetry, let us digress a little and
consider the equivariant index or refined index one encounters
in literature on wall-crossing, of the generic form
\begin{equation}
{\rm Tr}\left((-1)^Fy^{2 j_3}\right)
\end{equation}
with a ``rotation" generator $ j_3$ along $z$-axis. Most such
computations are based on some version of low energy quantum mechanics
on the classical moduli spaces, our $\CM_n$'s, but as we saw above,
the ``rotational symmetry" of $\CM_n$ is in fact not the purely
spatial rotation but a diagonal subgroup of spatial rotation $SU(2)_L$
and the field theory R-symmetry $SU(2)_R$. Therefore, the refined indices
that have been computed are in fact
\begin{equation}
{\rm Tr}\left((-1)^Fy^{2 \CJ_3}\right)= {\rm Tr}\left((-1)^Fy^{2 J_3+2I_3}\right)
\end{equation}
so actually would equal the protected spin character
\begin{equation}
{\bf Tr}\left((-1)^{2J_3}y^{2 J_3+2I_3}\right)
\end{equation}
of the field theory, if we are allowed to choose the
chirality operator $(-1)^F$ of the quantum mechanics to be $(-1)^{2J_3}$.

So this brings the question of what happens to the two natural chirality
operators, $(-1)^{2J_3}$ and $(-1)^{2I_3}$, when we deform and reduce the
dynamics in favor of a $\CM_n$ nonlinear sigma-model. As we saw, the two
$SU(2)$ symmetries are lost individually, so operators like $J_3$ and $I_3$
can no longer be used to classify eigenstates. Nevertheless,
$(-1)^{2J_3}$ and $(-1)^{2I_3}$ are still sensible chirality
operators.  Even after the deformation,  one can show
directly $(-1)^{2I_3}$ as a product of all fermions  while $(-1)^{2J_3}$ is
again the same product of all fermions times $(-1)^{\sum_{A>B} \langle
\g_A,\g_B\rangle +n-1}$. Both anticommute with the surviving supercharge $Q_4$,
so still defines chirality operators. This is not much of surprise since
they simply measure the most rudimentary information about the states,
i.e., whether, before deformation, the state was in a integral or in a
half-integral representations.

When we reduced the dynamics to $\CM_n$, however, we must properly redefine
these chirality operators by evaluating them on vacuum of the heavy oscillators.
For instance, consider $(-1)^{2I_3}$ for the simplest $n=2$ case. The canonical
chirality operator on $\CM_n=S^2$ is, as noted before,
$$(-1)^{F_{S^2}}=2i\hat \psi^1\hat\psi^2 \ ,$$
with the natural orientation arising from embedding of $S^2$ to $R^3$. To relate this to $(-1)^{2I_3}$,
we remember to  set the heavy fermions, $\psi^3 $ and $\lambda$, to their ground state,
which gives precisely
$$\langle0\vert(-1)^{2I_3}\vert0\rangle_{heavy}=(-1)^{F_{S^2}}\ , $$
it turns out.\footnote{ This can be seen most easily when
 we choose the ordering of $\g_A$'s such that
$\langle \g_A,\g_B\rangle$ are all nonnegative for $A>B$, which is
also the convention chosen in Ref.~\cite{Manschot:2010qz} for
non-scaling cases.} Clearly, we may repeat this for each sector of
$4$ fermions labeled by $A$, and find
\begin{equation}
\langle0\vert(-1)^{2I_3}\vert0\rangle_{heavy}= \prod_A 2i\hat \psi^1_A\hat\psi^2_A =(-1)^{F_{\CM_n}}\ .
\end{equation}
Therefore, the chirality operator $(-1)^{2I_3}$ prior to
the deformation, smoothly descend to the canonical chirality
operator on $(-1)^{F_{\CM_n}}$, upon deformation and subsequent
reduction of dynamics, and leads to the standard Dirac index $\CI_n$.

Since the desired index $\Omega$ needs $(-1)^{2J_3}$ as the chirality
operator, we then use (\ref{sign}) to relate $(-1)^{2J_3}$ to $(-1)^{2I_3}$,
and find an unambiguous answer,
\begin{equation}
\Omega^{distinct}=(-1)^{\sum_{A>B}\langle\g_A,\g_B\rangle +n-1}\times \CI_n(\{\g_A\})\ .
\end{equation}
On the left hand side, we emphasized the fact we are yet to incorporate the statistics
issue. We will see in next section how this generalizes when we impose
statistics to the index computation.
Before asking the question of statistics, however, there is still one more
ambiguity to the expression above, since so far we did not take into
account of the internal degeneracy and quantum numbers of the individual
charge centers. The left hand side
is still defined with respect to $(-1)^{2J_3}$, so adding internal degeneracy
factor can be accomodated by writing
\begin{equation}
\Omega^{distinct}=(-1)^{\sum_{A>B}\langle\g_A,\g_B\rangle +n-1}\times \CI_n(\{\g_A\})\times\prod_{A=1}^n\Omega_A
\end{equation}
where individual $\Omega_A$'s are also computed as the trace of $(-1)^{2J_3}$
(as usual modulo the universal half-hypermultiplet part).
As usual, we assume that there is no significant coupling of these internal degeneracy
to the quantum mechanical degrees of freedom. Sometimes, we will
also write this as
\begin{equation}\label{signfixed}
(-\Omega^{distinct})\times(-1)^{\sum_{A>B}\langle
\g_A,\g_B\rangle}=\CI_n(\{\g_A\})\times \prod_A(-\Omega_A)\,
\end{equation}
which is more convenient when keeping track of statistics, since, for $SU(2)_R$
singlets, the Bose/Fermi statistics are naturally correlated with the
sign of $-\Omega_A$'s.

\section{Index with Bose/Fermi Statistics and Rational
Invariants $\bar\Omega$}

So far, we pretended that dyons involved are all distinct,
and studied the supersymmetric bound states thereof. In
reality, this is not quite good enough since we often need to
understand bound states of many identical dyons, obeying
either fermionic or bosonic statistics.
The effective true moduli space, for example, has to be
an orbifold of type
\begin{equation}
{\CM_n}/{\Gamma}
\end{equation}
where $\Gamma$ is a union of permutation groups that
mix up labels for identical particles, with proper
action on wavefunctions. Equivalently,
the index should be computed with appropriate projection
operator inserted,
\begin{equation}
\Omega={\rm Tr}\left((-1)^{2J_3}e^{-s H}{\cal
P}_\Gamma\right)
\end{equation}
where $\CP_\Gamma$ projects to wavefunctions obeying
either Bose or Fermi statistics under the exchange of
identical particles.

The  orbifolding reduces the volume of the moduli space,
so given the index formula which is an  integral over the manifold,
we should expect to see factors like $1/d!$ as a result of having $d$
identical centers. However, action of $\Gamma$
is not everywhere free, since when  identical
particles are on top of one another, the action is trivial.
There are complicated fixed submanifolds under $\Gamma$,
making the problem very involved, and in particular
there should be additional contributions from the fixed
manifolds under the orbifolding action.

\subsection{The MPS Formula}

Before we carry out such a computation directly, it is instructive
to recall a recent result by Manschot, Pioline, Sen (MPS),
who evaded this complication algebraically, and replaced it by a
sum of many index problems with distinct charge centers \cite{Manschot:2010qz}.
They argued that one can recover the correct index, by adding
indices for a series of artificial problems with a smaller number of
charge centers. In this set of effective index problems, the
trick requires the following rules: When the reduced problem
has $d$ particles of the same kind, MPS divides the index
by $1/d!$. When one has a particle of nonprimitive charge as
a part of such a reduced problem, one must also use,
in place of the true intrinsic degeneracies $\Omega$
of the particles, a mathematical one $\bar\Omega$,
\begin{equation}
\bar\Omega(\gamma)\equiv \sum_{p\vert \gamma}
\frac{\Omega^+(\gamma/p)}{p^2}\ ,
\end{equation}
where the sum is over the positive integer $p$ such that
$\g/p$ belongs to the quantized charge lattice of
the theory. Note that $\Omega(\g)=\bar\Omega(\g)$
whenever $\g$ is primitive and $\bar\Omega(p\g)=\Omega(\g)/p^2$
if no non-primitive charge state exists.

For illustration, let us take two primitive charges $\beta_1$ and $\beta_2$.
Suppose that, among all possible linear combinations of the two,
only these two states exist on one side of the
marginal stability wall. Labeling the degeneracy by $\pm$
depending on which side of the wall we are considering,
we thus assume that
\begin{equation}
\Omega^+(m\beta_1+k\beta_2) = 0, \quad {\rm unless}\; (m,k)=\pm
(1,0) \;{\rm or}\; (m,k)=\pm (0,1)\ .
\end{equation}
The sign of $\Omega_{1,2}\equiv \Omega^+(\beta_{1,2})$ are
correlated with the statistics
assignment of the particle; a hypermultiplet has $\Omega=1$
and must be treated as fermions while a vector multiplet
has $\Omega=-2$ and must be treated as bosons. Under this
assumption, we have
$\bar\Omega(p\beta_{1,2})=\Omega(\beta_{1,2})/p^2$.
Manschot et.al.'s formula then simplifies to,
\begin{eqnarray}\label{simpleMPS}
&&-\Omega^-(m\beta_1+k\beta_2) \times (-1)^{\sum_{A>B}
\langle\g_A,\g_B\rangle}\nonumber\\
&=& \frac{1}{m!k!}
\,\CI_{m+k}(\beta_1,\beta_1,\dots,\beta_2,\beta_2,\cdots)
\,(-\Omega_1)^m(-\Omega_2)^k\nonumber\\
&+& \frac{1}{(m-2)!k!}
\,\CI_{m-1+k}(2\beta_1,\beta_1,\beta_1,\dots,
\beta_2,\beta_2,\dots) \,
\frac{-\Omega_1}{2^2}
\,(-\Omega_1)^{m-2}(-\Omega_2)^k\nonumber\\
&+& \frac{1}{(m-3)!k!}
\,\CI_{m-2+k}(3\beta_1,\beta_1,\beta_1,\dots,
\beta_2,\beta_2,\dots) \,
\frac{-\Omega_1}{3^2}
\,(-\Omega_1)^{m-3}(-\Omega_2)^k\nonumber\\
&+& \frac{1}{2!(m-4)!k!}
\,\CI_{m-2+k}(2\beta_1,2\beta_1,\beta_1,\beta_1\dots,
\beta_2,\beta_2,\dots) \,
\left( \frac{-\Omega_1}{2^2} \right)^2
\,(-\Omega_1)^{m-4}(-\Omega_2)^k\nonumber\\ \nonumber\\
&+&\cdots\ ,
\end{eqnarray}
where the sum is over all unordered partitions of $m\beta_1+k\beta_2$
respectively, although we listed above only part of the partitions of $m$.
For the overall sign, we re-labeled the individual charges
$\beta_1,\beta_1,\dots,\beta_2,\beta_2,\dots$ and called them $\g_A$'s.
This sum and each term in it can be characterized by the following set of rules:
\begin{enumerate}

\item[(i)] The sum is over all unordered partition of
$m\beta_1+k\beta_2=\sum_s d_s \beta_s$
where $\beta_s=(p_{s1}\beta_1+p_{s2}\beta_2)$. For each
$\beta_s$, we will have a factor $\bar\Omega(\beta_s)$,
so we can, with the current assumption on $\Omega^+$'s, consider only a
subset where only one of $p_{s1}$ and $p_{s2}$ is nonzero for each $s$.

\item[(ii)] The index $\CI_{n'}$ with ${n'\equiv \sum_s d_s }$
effective charge centers.
For $\CI_{n'}$, we treat all charge centers as distinguishable,
so it is computed by the index theorem of the previous section with $n'\le n$
distinguishable centers.

\item[(iii)] The combinatoric factor of $1/d_s!$ for each $s$.
This takes into account of the reduced volume of the moduli space due to the orbifolding
by the permutation subgroup $S(d_s)$ acting on the reduced
$n'$-center quantum mechanics, but does not address the
contribution from the submanifolds fixed by  $S(d_s)$.

\item[(iv)] For each effective particle of charge $p\beta $,
with primitive $\beta$ and $p>1$,
that shows up in computation of $\CI_{n'}$, one further assigns an
effective internal degeneracy factor
$-\bar\Omega(p\beta)=-\Omega(\beta)/p^2$, in addition to
$(-\Omega_1)^{m'}(-\Omega_2)^{k'}$, which reflects the fact that
$m'$ number of $\beta_1$ centers and $k'$ number of
$\beta_2$ centers are left as individual.
\end{enumerate}

The last $-\bar\Omega(p\beta)=-\Omega(\beta)/p^2$ should be compared to
the naive $(-\Omega(\beta))^p$ degeneracy factor that would be the correct
factor if we were considering $p$ separable particles of
charge $\beta$ instead of one particle of charge $p\beta$.
Finally, the appearance of $-\Omega$'s instead of $\Omega$'s is natural,
since for example a half-hypermultiplet with $\Omega=1$ acts like fermions,
while a vector multiplet with $\Omega=-2$ acts like bosons.

{}From the quantum mechanics viewpoint, the decomposition (i) clearly has
something do with the orbifolding action $\Gamma$. Each term in (\ref{simpleMPS}) arises from a
submanifold which is fixed by the product of permutation groups of order $p_s$,
$\prod_s S(p_s) \subset \G$. For each sector, origins of (ii) and (iii) are
also evident as coming from a reduced problem of $n'$ charge centers and the
subsequent volume-reducing action of $\prod_s S(d_s) = \G/\prod_s S(p_s) $.
The only part of this formula which is not evident, so far, from the moduli quantum mechanics
viewpoint is the rational degeneracy factor of (iv). Here,
we would like to isolate where this comes from, and later
derive it directly from the moduli dynamics.

After some careful thinking, it becomes evident that this rational degeneracy
factor should come from quantum mechanical degrees of freedom normal to the
submanifold fixed by $S(p)$'s. Let us consider only $p>1$ cases and label them $p_{s'}$,
since otherwise the internal degeneracy factor is $\Omega(\beta)$ as expected.
Subgroup $S(p_{s'})$'s permuting these $p_{s'}>1$ charges
fixes a submanifold $\CM_{n-\sum (p_{s'}-1)}$ inside $\CM_n$.
This fixed submanifold has a codimension $2\sum(p_{s'}-1)$ in $\CM_n$, since it
is spanned by coincidence of $p_{s'}$ centers, each of which span two
directions in $\CM_n$.

Consider the reduced dynamics on the intersection,
$\CM_{n'=n-\sum (p_{s'}-1)}$, for computation of $\CI_{n'}$ with
all such $p_{s'}\beta_{s'}$ center treated as single particle,
respectively. If we start with this reduced index
problem, impose the statistics, and pretend that the centers associated with
$p_{s'}\beta_{s'}$ comess with a unit degeneracy we will find a contribution of type
\begin{equation}
\frac{1}{\prod_s d_s!}\ \CI_{n'}\times
(-\Omega_1)^{m'}(-\Omega_2)^{k'}\ ,
\end{equation}
where
$m'\beta_1+k'\beta_2=m\beta_1+k\beta_2-\sum_{s'}p_{s'}\beta_{s'}$
and $n'=m'+k'+\sum_{s'}d_{s'}$.
Note that we took care to include the volume-reducing effect of
$\prod_s S(d_s)=\G/\prod S(p_{s'})$
via the denominator  $\prod d_s!=m'!k'! ( \prod d_{s'}!)$.

This expression is obtained after ignoring the quantum degrees of freedom
that are normal to the fixed manifold $\CM_{n-\sum (p_{s'}-1)}$'s, and
does not agree with MPS formula. The latter is
\begin{equation}
\frac{1}{\prod_s d_s!}\; \CI_{n'} \times (-
\Omega_1)^{m'}(-\Omega_2)^{k'}
\times \prod_{s'}\frac{-\Omega(\beta_{s'})}{p_{s'}^2}
\end{equation}
so the difference is precisely the rational degeneracy factor of (iv).
Clearly it comes from quantizing the normal bundles
of $\CM_{n-\sum (p_{s'}-1)}$'s inside $\CM_n$. On the other hand,
for all intent and purpose, this part of quantum mechanics is
free, since they have something to do with many identical particles
and has no interaction of type $\CL_1$, except for the
statistics issue.\footnote{There is a subtlety,
again related to whether threshold bound state of identical charges
can form. Since we started with the assumption that such nonprimitive
state do not exist, it is safe to assume this issue does not
complicate our problem. Whether or not we can extend this to
theories with threshold bound states, i.e. supergravity, is an
open problem.}

This leads us to conclude that the factor, $\Omega(\beta)/p^2$,
should arise from an index of $p$ noninteracting identical particles
of charge $\beta$, modulo
the center of mass part which already contributed to $\CI_{n'=n-p+1}$.
The relative dynamics of such identical particles carry $2(p-1)$ bosonic degrees of freedom, $2(p-1)$ fermionic
degrees of freedom, and additionally internal degeneracy of $|\Omega(\beta)|$
for all $p$ particles. In next subsection, we will show that precisely
such a factor arises from the dynamics of non-interacting and identical $p$
particles with the internal degeneracy $\Omega(\beta)$.

The full MPS formula follows the same set of rules, except that
one must in general consider an arbitrary set of charges on the + side, and all
the partitions of the total charge $\gamma_T$ in terms of charges of
states that exist on $+$ side of the wall.
Since the + side of spectrum may then include states of
charges $h\beta_1+j\beta_2$ with $h+j>1$, more diverse
charge centers will appear for the individual index problems on the right hand side.
As we will discuss later, this can be incorporated by treating
all such particles on the + side as independent. The
only subtlety is when non-primitively charged states
exist on the + side; this can be remedied by employing
the fully general form,
$
\bar\Omega(\gamma)\equiv \sum_{p\vert \gamma}
{\Omega^+(\gamma/p)}/{p^2}
$ as the effective degeneracy factor.
We will also see this most general $\bar\Omega$ emerging from our index
computations.

\subsection{Physical Origin of $\Omega(\beta)/p^2$ from $p$
Non-Interacting Identical Particles}

Let us first restrict ourselves to bound states involving several identical
dyons of charge $\beta$ with $-\Omega^+(\beta)=\pm 1$.\footnote{Because of
spin-statistics theorem, there is no irreducible BPS multiplet in $D=4$ $N=2$ theories with
$\Omega=-1$. The half-hypermultiplet has 1 and vector multiplet has -2.
The assumption here is strictly for the illustrative purpose only.}
As in the previous discussion, let us consider the bound state of
$n$ charge centers, $\g_T=\sum_A\g_A$, $m$ of which are $\beta$'s.
The identical nature of the $\beta$ dyons means that the orbifolding
group includes the $m$-th order permutation group $S(m)$.
We start with the assumption, for simplicity, that
$k\beta$ state exists only for $|k|=1$ on the + side,
 and then come back for the fully general case in next subsection.

Consider the index reduced on the true moduli space $\CM_{n}$
as described in the previous section, with proper account taken of
the bosonic or the fermionic statistics,
\begin{equation}\label{true}
\Omega^-\left(\sum_A\g_A\right)-\Omega^+\left(\sum_A\g_A\right)
=\int_{\CM_n}{\rm tr}\left(\langle X|(-1)^{2J_3}e^{-s
Q^2}\CP_\Gamma|X\rangle\right)\,dX\ ,
\end{equation}
via the orbifolding projection operator $\CP_\Gamma$.
Here ${\rm tr}$ means the trace over the fermionic variables as well
as other internal discrete degrees of freedom, and
we integrate over the bosonic variables $X$ with an appropriate measure.
Matching the sign of $-\Omega$ with $(-1)^F$ value of the component
dyon states, as we noted in the case of bound state counting in
distinguishable centers, this index naturally computes the degeneracy
$\Omega^-$'s as
\begin{equation}
{\rm Ind}(\{\g_A\};\Gamma) =  \int_{\CM_n}{\rm tr}\left(\langle X|(-1)^{2J_3}e^{-s
Q^2}\CP_\Gamma|X\rangle\right)\,dX\ ,
\end{equation}
so we would like to ask whether this reproduces
(\ref{simpleMPS}) and rediscover the rational
invariant $\bar\Omega$. For this, let us concentrate on
the permutation group $S(m)$ part of $\Gamma$ and see how it
generates a series of terms, similar to MPS's wall-crossing formula.

Inside $\CM_n$ there are various fixed submanifolds, $\CM_{n'}$, of
dimension $2(n'-1)$. The simplest are $\CM_{n-p+1}$, fixed under
$S(p)$ subgroup of $S(m)$.
Note that we use the same notation $\CM$ for the
fixed submanifolds as the full classical moduli manifold $\CM$.
This is because all of them are of exactly the same type. For example,
the manifold $\CM_{n-m+1}$ would also emerge if we started with a different
low energy dynamics involving a single center of charge $m\beta$ in place
of $m$ centers of charge $\beta$. Ignoring contributions from these
fixed manifolds would simply give
\begin{equation}
(-1)^{\sum_{A>B} \langle\g_A,\g_B\rangle+n-1}\left(\frac{\CI_n}{|\Gamma|} \right)\times \prod_A\Omega_A
\end{equation}
due to the volume-reducing action of $\Gamma$ when it is acting freely.
This is the very first term in the MPS formula. Since there are many
fixed submanifolds, however, each of them will contribute  additively
on top of this.

Without loss of generality, let us consider the fixed manifold $\CM_{n-p+1}$
associated with the partition $m \beta = p\beta+\beta+\beta+\cdots+\beta$,
and label the coordinates along the fixed manifold $\CM_{n-p+1}$
by $X'$ and those normal to it by $Y$.
Note that among $X'$ are the two coincident (or center of mass) coordinates
for the $p\beta$ charge center, so we can think of $Y$'s as the relative
position coordinates among these $p$ charge centers; therefore there are
$2(p-1)$ $Y$'s.
We then formally write the additive contribution from the
fixed submanifold $\CM_{n-p+1}$ as
\begin{eqnarray}
&&\Delta_p\times {\rm Ind}(\{\g_{A'}\}=\{p\beta,\beta,\dots\};\Gamma')\cr\cr
&=&\Delta_p\times\int_{ \CM_{n-p+1} }{\rm tr}'\left(\langle Y=0,
X'|(-1)^{2J_3'}
e^{-s H'}\CP_{\Gamma'}|Y=0, X'\rangle\right)\,dX'\ ,
\end{eqnarray}
where $\G'=\G/S(p)$ is the remaining orbifolding group that
acts nontrivially on $\CM_{n-p+1}$. Here tr$'$ denotes trace over
fermionic and other internal degrees of freedom, except those associated
with the $p$ identical $\beta$'s that are held together at $\CM_{n-p+1}$.

We factored out the contribution  $\Delta_p$ from the normal directions,
$Y$, and the superpartners thereof. On the other hand,
the second factor is the index of a reduced $n-p+1$ center problem,
modulo the internal degeneracy factor of $p\beta$ charge center. Other than this, the computation of this
latter factor proceeds on equal footing as (\ref{true}),
\begin{eqnarray}
&&\int_{ \CM_{n-p+1} }{\rm tr}'\left(\langle Y=0,X'|(-1)^{2J_3'}
e^{-s H'}\CP_{\Gamma'}|Y=0, X'\rangle\right)\,dX'\cr\cr
&&\simeq
(-1)^{\sum_{A'>B'} \langle\g_{A'},\g_{B'}\rangle+n-p}
\left(\frac{\CI_{n-p+1}(\{\g_{A'}\})}{|\Gamma'|} \right)\times \prod_{A'=2}^{n-p+1}\Omega_{A'}+\cdots
\end{eqnarray}
so we may compute the full index recursively.  $\Delta_p$ plays the
role of the missing internal degeneracy factor $\Omega_{1'}$ here, as it
computes the effective contribution from these $p$ coincident $\beta$'s.
The ellipsis denotes terms from other fixed submanifold inside $\CM_{n-p+1}$ etc.

We will show that $\Delta_p=\pm 1 /p^2$, regardless of precise
nature of the $\beta$ particles, which also reproduces MPS formula
for $\Omega(\beta)=\pm 1 $ entirely from the dynamics. Schematically,
this factor can be written as
\begin{equation}
\sim
\int {\rm tr}\left(\langle Y|(-1)^{2J_3^\perp}\,e^{-s H^\perp
}{\mathcal{P}}|Y\rangle\right) dY\ ,
\end{equation}
with $F_\perp$ and $H_\perp$ defined on $Y$'s and the superpartners,
again with suitable measure for the bosonic integral. The projection operator
\begin{eqnarray}
\mathcal{P}=\frac1{p\,!}\sum_{\pi\in S(p)}(\mp
1)^{\sigma(\pi)}M_{\pi}
\end{eqnarray}
ensures that we isolate wavefunctions of correct Bose/Fermi
statistics. Note that the sign in the projection operator
is the same as that of $-\Omega(\beta)$. $M_\pi$ is the $(p-1)$ dimensional
representation of  $\pi \in S(p)$, common for $(p-1)$
coordinate doublets $Y's$ and  for their fermionic partners,  $\psi$'s.
Naturally ${\sigma(\pi)}$ is odd or even when $\pi$ is odd or even.

Since the embedding of $\CM_{n-p+1}$'s in $\CM_n$ could be very
complicated, the exact nature of the decomposition is not entirely clear.
On the other hand, the initial index problem is gapped and allows us
to take $s\rightarrow 0$. At least in this limit, the decomposition
makes sense intuitively, and, as we will see shortly it suffices to
consider an arbitrarily small tubular neighborhood around the fixed
manifold $\CM_{n-p+1}$. We take the $Y$ directions as
 a ball $B_{2(p-1)}$ insides a flat $R^{2(p-1)}$. Therefore, we have
\begin{equation}
\Delta_p = \lim_{s\rightarrow 0}\int_{B_{2(p-1)}} {\rm
tr}\left(\langle Y|\,(-1)^{J_3^\perp}
e^{(s/2) \nabla^2}{\mathcal{P}}|Y\rangle\right) dY \ .
\end{equation}
Precisely how we cut-off this neighborhood will not matter,
as we will see shortly that a Gaussian
integral of squared  width $s$ emerges along $Y$ directions.

Interestingly, exactly the same kind of object was studied when
solving for the famous D0 bound state problems in the 1990's. The first
such computation appeared in Ref.~\cite{Yi:1997eg} on two-body problem and was
later expanded to many body case in Ref.~\cite{Green:1997tn}. Here
we will adapt and expand the computations in these works. Since there are
$2(p-1)$ real fermions,
we will choose a polarization of type $\{\psi_i,\psi^\dagger_i
\}$, so that a general wave function  $\mid \Psi\rangle$ can be expanded as
\begin{eqnarray}
|\Psi\rangle = \left(\Psi(Y)+\Psi^{\{i\}}(Y)\psi^\dagger_{(i)}
+{1
\over2}\Psi^{\{i_1i_2\}}(Y)\psi^\dagger_{(i_1)}\psi^\dagger_{(i_2)}
+\dots\right)|0\rangle\ ,
\end{eqnarray}
where $\Psi^{\{i_1\dots i_m\}}(Y)=\sum_k\lambda^{\{i_1\dots
i_m\}}_k\Psi_{k}(Y)$
and $\{\Psi_{k}\}$ are complete basis of $Y$-space wave functions.
Since the Hamiltonian is free and does not mix sectors of different
fermion numbers, we may evaluate the bosonic and fermionic trace
independently. The fermionic trace, for each of $M_\pi$, is given by
\begin{eqnarray}
&& (\pm 1)^p\,{\rm
tr}_{\psi}\left((-1)^{p-1}(-1)^{N_{\psi^\dagger}}M_{\pi}\right)\ ,
\end{eqnarray}
where $(\pm 1)^p$ arises from the value of $(-1)^{2J_3}$ on $p$
individual $\beta$ states. Also the orbital part of $J_3^\perp$ is
always integral in  the absence of the minimal coupling contribution,
so $(-1)^{2J_3^\perp}$ acting on the quantum mechanical degrees of
freedom becomes purely  fermionic expression $(-1)^{p-1}(-1)^{N_{\psi^\dagger}}$.
The sign in front of the latter comes from converting the chirality
operator to a form involving the fermion number operator that counts
the creation operators $\psi^\dagger$.

Then the contribution from $Y$ direction reads
\begin{eqnarray}
\Delta_p
&=&\frac{(\pm 1)^p}{p\, !}\;\sum_{\pi\in S(p)}(\mp 1)^{\sigma_{\pi}}(-1)^{p-1}
{\rm tr}_{\psi}\left((-1)^{N_{\psi^\dagger}}M_{\pi}\right)\cr
&&\hskip 2.5cm \times \int_{B_{2(p-1)}} \langle
Y|\,e^{s\nabla^2/2}M_{\pi}|Y\rangle dY \ ,
\end{eqnarray}
in the limit of $s\rightarrow 0$.
A crucial observation that allows us to proceed systematically is
\begin{eqnarray}
&&{\rm tr}_{\psi}\left((-1)^{F^\perp}M_{\pi}\right) \cr
&=&
\langle 0|0\rangle - \langle 0| \psi^{a'} M_{\pi
a'}{}^{a}\psi^\dagger_a| 0\rangle
+{1\over 2}\langle 0|\psi^{b'}\psi^{a'}M_{\pi a'}{}^{a}M_{\pi
b'}{}^{ b}\psi^\dagger_a\psi^\dagger_b|0\rangle
+\cdots\cr
\cr&=&{{\rm det}(1-M_{\pi})}\ ,
\end{eqnarray}
and furthermore
\begin{eqnarray}
{\rm det}(1-M_{\pi})&=&\left\{
\begin{array}{ccc}
  p & ,~ & \pi \hbox{ is a cyclic permutation of order $p$ } \\
  0 & ,~ & \hbox{otherwise}
\end{array}\right.
\end{eqnarray}
for which it is important to remember that $M_\pi$ is a $p-1$ (rather than $p$)
dimensional representation of $S(p)$.
Since $(-1)^{\sigma_\pi}=(-1)^{p-1}$ for any cyclic permutation
of order $p$, we find
\begin{eqnarray}
\Delta_p
&=&
\frac{(\pm 1)^p}{p\,!}\;\sum_{\pi'}(\mp 1)^{p-1}(-1)^{p-1} {\rm
det}(1-M_{\pi'})
\int \langle Y|\,e^{s\nabla^2/2}M_{\pi'}|Y\rangle dY \cr
&=&
\frac{\pm 1}{p\,!}\;\sum_{\pi' }{\rm det}(1-M_{\pi'})
\times \frac{1}{(2\pi s)^{p-1}}\int e^{-(Y-M_{\pi'} Y)^2/2 s} dY\cr
&=&
\frac{\pm 1}{p\, !}\;\sum_{\pi'}\frac{1}{{\rm
det}(1-M_{\pi'})} \ ,
\end{eqnarray}
where the sum is now only over the cyclic permutations of order
$p$. There are precisely $(p-1)!$ such permutations and they each
contribute
$1/p$, via the determinant, so the result is
\begin{equation}
\Delta_p=\frac{\pm 1}{p^2} \ ,
\end{equation}
as promised. Clearly, we can repeat this when there are several
such factors simultaneously, to give,
\begin{equation}
\Delta_{\{p_{s'}\}}=\prod_{s'} \frac{\pm 1}{p_{s'}^2} \ ,
\end{equation}
reproducing $\Omega(\beta)/p_s^2$  of the MPS formula
with $\Omega(\beta)=\pm 1$.

The more general case of $\Omega(\beta)=\pm d$  can be derived
similarly. Let us write one particle state as
\begin{eqnarray}
|\hat\Psi\rangle=\hat\Psi^{\eta}(x,\psi)|0;\eta\rangle
\end{eqnarray}
so that $\eta=1,2,\dots,d$ labels the internal degeneracy,
and  $p$-particles wave function (without center
of mass degree of freedom) can be
written as a sum of terms like
\begin{eqnarray}
\Psi^{\{ij\cdots
k\}\{\eta_s\}}(Y)\psi^\dagger_{(i)}\psi^\dagger_{(j)}\cdots\psi^\dagger_{(k)}
|0;\eta_1,\eta_2,\dots,\eta_p\rangle\,,
\end{eqnarray}
none of which mixes under the free Hamiltonian. Thanks to this,
Just as the fermionic part and the bosonic part separately contributed,
this internal part also factorizes under each $\pi$. Expressing
$\Delta_p$ as a sum over the elements of permutation group
again, we now have an extra factor
$$\langle\eta_p,\dots,\eta_1|\eta_{\pi(1)},\dots,\eta_{\pi(p)}\rangle\ ,$$
for each permutation $\pi$, and the trace over these internal indices.
Thus, we arrive at a similarly simple form,
\begin{eqnarray}
\Delta_p=
\frac{(\pm 1)}{p\,!}\;\sum_{\eta_1,\dots,\eta_p}\sum_{\pi'}
\frac{\langle\eta_p,\dots,\eta_1|\eta_{\pi(1)},\dots,\eta_{\pi(p)}\rangle}{{\rm
det}(1-M_{\pi'})} \ .
\end{eqnarray}
As before, the sum is over the permutations of cyclic order $p$.
For such $\pi'$, the inner product vanishes identically unless all
$\eta_i$'s are equal to one another, and gives unit if all are equal.
The sum over $\eta$'s thus collapses to a single sum over
$\eta=\eta_1=\eta_2=\cdots=\eta_p$, and
\begin{eqnarray}
\Delta_p=(\pm 1)\sum_\eta 1\times \frac{1}{p\,!}\sum_{\pi'}
\frac{1}{{\rm det}(1-M_{\pi'})}=\frac{\pm
d}{p^2}=\frac{\Omega(\beta)}{p^2} \ .
\end{eqnarray}
This gives us the only essential ingredient in confirming
(\ref{true}), and in fact the fully general version thereof.

For a complete derivation, a recursive argument is needed
and we need to consider possibility of low energy dynamics with
charge centers both primitive and non-primitive charges
simultaneously. This naturally brings us to the most general
wall-crossing formula, next.

\subsection{General Wall-Crossing Formula}

Most of what we derived generalizes to cases with arbitrary spectrum on + side,
 without much modification, but here we need to point out one subtlety.
Suppose the $+$ side of spectrum contains not only a pair of primitively charge
states $\g$ and $\g'$
but also states like $h\g+ j \g'$ a little more involved,
and includes states of composite charges such as $m\g$ or linear combinations
with other charges. (One can also have states with charges completely
unrelated to these  but those will not participate in the wall-crossing,
and therefore irrelevant.)
Such a state cannot be considered as a bound state of $h$ $\g$'s and $j$ $\g'$'s,
since the two are mutually repulsive on the $+$ side.
Rather, it should be regarded as a completely independent
particle of different origin. In fact, for $SU(2)$ theory with a single flavor,
a monopole $\g$, a quark $\g'$ and a dyon $\g+\g'$ are known to coexist
in the central part of the moduli space.

Let us denote charges of these independent particles as $\beta_v$.
Since one can form bound states of a given total charge
$\g_T$ on the ``$-$" side from different combinations of these $+$ side
states. We will label each of these physically distinct combination by the
upper index inside a parenthesis such that
\begin{eqnarray}
\g_T =\sum_A m_v^{(1)}\beta_v =\sum_v m_v^{(2)}\beta_v=\sum_v
m_v^{(3)}\beta_v=\cdots\ .
\end{eqnarray}
In such circumstances, the total degeneracy for $\g_T$ has to
be computed for each of such bound state problems and summed over,
\begin{equation}
\Omega^-(\g_T)-\Omega^+(\g_T)=\sum_q
\Omega\left(\{m_v^{(q)}\}\right)\ ,
\end{equation}
where each term on the right hand side is computed from the
$n^{(q)}=\sum m_A^{(q)}$ center quantum mechanics.
For each of $\Omega(\{m_v^{(q)}\})$'s, computation of the
previous subsection goes
through without modification, and will be computed as
\begin{eqnarray}
\Omega\left(\{m_v^{(q)}\}\right)&=&\Omega\left(\sum_A\g_A
=\sum_v
m_v^{(q)}\beta_v\right)\ .
\end{eqnarray}
One important detail to remember is that, even if
some of charge $\beta_v$'s might be a linear combination of other $\beta_v$'s, each
of them are physically unrelated independent  particles: The permutation
group is simply $\Gamma=\prod_v S(m^{(q)}_v)$ for each of these index problems.

Combining this with the results of previous subsection, we reproduce MPS
formula in its most general form.
Note that, when we reorganize this formula in terms of the index of distinct
particles of unit individual degeneracy,
\begin{equation}
\CI_n(\{\dots,\g,\dots\})=\int_{{\cal M}_n} Ch(\CF)\hat A(\CM_n)
\end{equation}
the rational invariants $\bar \Omega(\gamma)$, multiplying them,
will accumulate additive contributions of the form, including $p=1$ case,
\begin{equation}
\frac{\Omega^+(\gamma/p)}{p^2}
\end{equation}
for each $\gamma/p=\beta_v$ that appears in one of the expansion
$\gamma_T=\sum_v m^{(q)}_v\beta_v$. Since we are summing
over all possible such expansions, it implies that
\begin{equation}
\bar\Omega(\gamma)=\sum_{p\vert
\gamma}\frac{\Omega^+(\gamma/p)}{p^2}
\end{equation}
will appear as the effective degeneracy factors that
multiply $\CI_n$'s. $p=1$ terms arises only when $\gamma$ is
one of the $\beta_v$'s, while $p>1$ terms arise from the
orbifold fixed sector as in the previous section when
$\gamma/p$ is one of  $\beta_v$'s. The final expression is
\begin{eqnarray}
&&\cr
&&\Omega^-(\g_T)-\Omega^+(\g_T) =\cr\cr
&&\cdots \cr\cr
&&+(-1)^{-n'+1+\sum_{A'>B'} \langle\g'_{A'},\g'_{B'}\rangle}
\times \frac{\CI_{n'}(\{\g'_{1'},\g'_{2'},\dots\})}{|\Gamma'|}
\times\prod_{A'}\bar\Omega(\g'_{A'})\cr\cr
&&+\cdots\\ &&\nonumber
\end{eqnarray}
where we wrote a representative form for the partition
$\gamma_T=\sum_{A'=1}^{n'}\g'_{A'}$ into $n'$ centers and the
associated orbifolding group $\Gamma'$ permuting
among identical elements  in $\{\g'_{A'}\}$.

\section{Alternative Derivations for  Arbitrary Number of Centers}

We have so far seen how Dirac index computes wall-crossing formulae
with the statistics imposed as a projection operator
in the index problem. In particular by decomposing the
computation into various fixed submanifold, we effectively
reduce the problem to a collection of index problems for
certain sets of distinguishable particles. Some of these
distinguishable particles originate from real BPS particles,
whereas some are mathematical construct. In the latter case, say
associated with any fixed submanifolds under the permutation
group $S(p)$, an effective internal degeneracy $\Omega(\beta)/p^2$
emerges universally.

In this section, we  repeat this exercise for general cases,
starting with  a string theoretical method of computing orbifold
index. While the derivation of previous section is very intuitive
and compelling, we wish to further support it with this more
rigorous formulation. To handle this problem, we have to figure out
how to define the index problem on orbifold singularities. This is a
subtle issue in mathematics since most index problems are defined on
smooth manifolds. On the other hand, in string theory the orbifold
singularities are common and the string theory on such space is well
defined. We will use the index for the Dirac-type operator defined
in the string theory setting and apply it to the problem of our
interest.

In fact in the original paper on the orbifold \cite{orbifold},
there appeared the formula of the Euler
characteristic on orbifolds. The special feature of this formula is
that it has the non-zero contribution from the twisted sectors and
it gives the right Euler-characteristic for the blown-up manifold
out of the underlying orbifolds when the way of the blown-up of
the orbifolds is known.
This consequently motivates many mathematical literatures which try
to justify the formulae in a rigorous way \cite{farsi}. For our
purpose, the natural object is the Dirac-Ramond index actively
studied on 1980's and we try to read off the index of interest.
We are interested in the $U(1)$ equivariant index since we will
follow closely the strategy developed by \cite{Manschot:2010qz}. It turns
out that the twisted sectors would not contribute and the entire
problem effectively reduces to the usual Atiyah-Bott-Lefshetz
theorem \cite{Atiyah}, with string theory serving as a computational
tool.

\subsection{Dirac-Ramond Operator on Orbifolds}

Now let us turn into the Dirac-Ramond index. Dirac-Ramond index
is so called the U(1) character-valued index \cite{Killingback}
\begin{equation}
I=\lim_{\tau_2\rightarrow 0} {\rm Tr} (-1)^{F_R} exp (2\pi i\tau_1
P) exp (-2\pi \tau_2 H)\ ,
\end{equation}
where $H$ is the Hamiltonian for the string worldsheet and $P$ is
the momentum of the string. Here $P$ plays the role of the $U(1)$
generator, which is the translation generator along the worldsheet.
With $\tau=\tau_1+i\tau_2$ and $q=e^{2\pi i\tau}$, this index has
the contribution from supersymmetric states of world-sheet only
and\footnote{We change the role of $q,\bar{q}$ in Eq. (\ref{ieq})
for later notational convenience.}
\begin{equation}
I={\rm Tr} (-1)^{F_R}
\bar{q}^{L_0+\epsilon}q^{\bar{L}_0+\bar{\epsilon}}={\rm Tr}
(-1)^{F_R}q^{-P} \ , \label{ieq}
\end{equation}
so that we have the contributions only from left-moving modes.
We define $H=L_0+\bar{L}_0, P=L_0-\bar{L}_0$ and $\epsilon,
\bar{\epsilon}$ is the zero point contributions from the right and
the left movers respectively. Thus $I$ has the form
\begin{equation}
I=\sum_{\lambda} \CI_{\lambda} q^{-\lambda}\ ,
\end{equation}
where $\CI_{\lambda}$ is the index on the subspace with momentum
$\lambda$, which is essentially the level of the massive string
state denoted by $m$. If we take the lowest $\lambda$ this will give
the usual index of the Dirac operator. $I$ has the form
\begin{equation}
I=\hat{A}(R)\sum_{(g,t,m)} q^m Ch(F,g) Ch(R,t)\ ,
\end{equation}
with $g$ representation of the gauge group and $t$ tensor
representation of the rotation group. The sum is taken over all
fields in representation $(g,t)$ at each mass level $m$ of the left
mover \cite{Warner, Wittenelliptic}.

Now $I$ is closely related to the string partition function. For
example, if we evaluate the Dirac-Ramond index for the NS fermions
with anti-periodic boundary conditions along worldsheet time
direction in the heterotic string theory, we have \cite{Li},
\begin{equation}
I= q^{-\frac{k}{24}-\frac{d}{24}}\int_M dx^{\mu}_0
d\psi^{\mu}_{-0}
\hat{A}(M) \prod_{n=1}^{\infty} \frac{det (1+q^{n-\frac{1}{2}}
exp\frac{iF}{2\pi})}{det (1-q^{n} exp \frac{iR}{2\pi})}\ ,
\end{equation}
where $x^{\mu}_0, \psi^{\mu}_{-0}$ denote the bosonic and fermionic
zero modes. When we set $F=R=0$ we have the typical form of the
string partition function. In the above we defined
\begin{eqnarray}
R_{\mu\nu}&=& \frac{1}{2}R_{\mu\nu\lambda\sigma}(x_0)
\psi^{\lambda}_{-0}\psi^{\sigma}_{-0} \ ,\nonumber \\
F_{AB}&=& -\frac{i}{2}F_{\mu\nu}^M T^M_{AB}
\psi^{\mu}_{-0}\psi^{\nu}_{-0}\ ,
\end{eqnarray}
and regard the fermion zero modes as the differential forms in the
usual way. For our purpose, the gauge bundle is Abelian, so
we simply replace $F$ here by ${\cal F}$ at the end of the
computation.

If we take the lowest $q$ this picks up the basic representation and $I$ gives
\begin{equation}
\int_M dx^{\mu}_0 d\psi^{\mu}_{-0} \hat{A}(M) Ch(F) \ ,
\end{equation}
where we use \cite{Warner}
\begin{eqnarray}
\sum x^k Ch (F, [k])&=&det (1+x exp \frac{iF}{2\pi}) \ ,\nonumber\\
\sum x^k Ch (F, (k))&=&det (1-x exp \frac{iF}{2\pi})^{-1}\ ,
\end{eqnarray}
with $[k], (k)$ being the antisymmetric and the symmetric products
of the basic representation. Thus in this way we obtain the formula
for the Dirac index from the Dirac-Ramond index.

Now this formalism can be easily generalized to the orbifold cases.
The twisted index of the Dirac-Ramond operator is given by \cite{Warner2}
\begin{equation}
I(g_1,g_2;\tau)={\rm Tr}_{g_2}
[\bar{q}^{L_0+\epsilon}q^{\bar{L}_0+\bar{\epsilon}} g_1\,
(-1)^{F_R}]
\end{equation}
where the $g_1, g_2$ are the twists along $\sigma_2, \sigma_1$
directions, respectively. Note that we define $g_1$ twists along
temporal direction. Here we label the points of the world sheet
torus by the complex quantity $\sigma_1+\tau \sigma_2$ with
$\sigma_1, \sigma_2$ having the periodicity $2\pi$. Consistency of
the boundary condition requires that $g_1, g_2$ commute. In
this case the bosonic and the fermionic zero modes exist only
along $M_{g_1g_2}$, which is the space inert under $g_1$ and
$g_2$. If we decompose the curvature along and the normal directions of
dimension $d_n$ to $M_{g_1g_2}$ whose dimension is $d$,
\begin{eqnarray}
\frac{1}{2\pi}R_{kl}=\frac{i}{4\pi}R_{kl\,
ij}\psi^i_{-0}\psi^j_{-0}&=& \left(
\begin{array}{cc}
                0 & \omega_{\mu} \\
-\omega_{\mu} & 0 \end{array} \right) \ ,\qquad p=1 \cdots \frac{d}{2}\ ,
\nonumber \\
\frac{1}{2\pi}R_{ab}=\frac{i}{4\pi}R_{ab\,
ij}\psi^i_{-0}\psi^j_{-0}&=& \left(
\begin{array}{cc}
                0 & \omega_p \\
-\omega_p & 0 \end{array} \right) \ , \qquad p=1 \cdots \frac{d_n}{2} \ ,
               \nonumber
               \end{eqnarray}
while gauge bundle of rank $d^E$ has the field strength
components
\begin{eqnarray}
\frac{1}{2\pi}F_{AB}=\frac{i}{4\pi}F_{AB\,
ij}\psi^i_{-0}\psi^j_{-0}&=& \left(
\begin{array}{cc}
                0 & \rho_r \\
-\rho_r & 0 \end{array} \right) \ , \qquad r=1 \cdots \frac{d^E}{2}\ ,
               \nonumber \\
               \end{eqnarray}
where $i,j,k,l$ denotes the indices along $M_{g_1g_2}$ directions.

Let's also define the action of $g_1, g_2$ on a real bundle $V$. $V$
could be either tangent bundle $TM$ or the gauge bundle $E$. Under
the action of $g_1,g_2$, $V$ restricted to $M_{g_1g_2}$ formally
decomposes as
\begin{equation}
V|_{M_{g_1g_2}}=V^{(0)}\oplus (\oplus_{r} V^{(r)}).
\end{equation}
The second term represents a formal sum of 2-dimensional real
bundles on which $g_1, g_2$ act as rotation by the angles $
 2\pi\alpha_r, 2\pi\beta_r$ respectively. Then,
\begin{eqnarray}
I(g_1,g_2;\tau)&=&\int_{M_{g_1g_2}}dx^i_0d\psi^i_{-0}
\frac{1}{\sqrt{det'
(\bar{\partial}+i\frac{R_{kl}}{2\pi})}} \nonumber \\
& & \sqrt{\frac{ \prod_{r=1}^{\frac{d^E}{2}}
det_{\alpha_r+\frac{1}{2}, \beta_r+\frac{1}{2}}
(\bar{\partial}+i\rho_r)det_{-\alpha_r-\frac{1}{2},-\beta_r-\frac{1}{2}}(\bar{\partial}-i\rho_r)}
{ \prod_{r=1}^{\frac{d_n}{2}} det_{\alpha_p, \beta_p}
(\bar{\partial}+i\omega_p)det_{-\alpha_p,-\beta_p}(\bar{\partial}-i\omega_p)}}\ ,
\end{eqnarray}
where $det'$ denotes the determinants without the zero modes. Here
we denote $det_{\alpha,\beta}(\bar{\partial}+\nu)$ denotes the
determinants of $\bar{\partial}+\nu$ of complex modes
$\Psi(\sigma_1, \sigma_2)$ with the boundary conditions,
\begin{eqnarray}
\Psi(\sigma_1+2\pi, \sigma_2)&=& e^{2\pi i
\beta}\Psi(\sigma_1,\sigma_2)\ ,
\nonumber \\
\Psi(\sigma_1, \sigma_2+2\pi)&=& e^{2\pi i
\alpha}\Psi(\sigma_1,\sigma_2)\ .
\end{eqnarray}
And
\begin{eqnarray}
& & det_{\alpha,\beta}(\bar{\partial})\equiv d
\left[
\begin{array}{c}
                \alpha  \\
\beta    \end{array}  \right] (0|\tau)\nonumber \\& &
=e^{i\pi(\alpha\beta-\beta)}q^{\frac{\beta^2-\beta+\frac{1}{6}}{2}}
\prod_{n=1}^{\infty}(1-q^{n-\beta}e^{2\pi
i\alpha})(1-q^{n+\beta-1}e^{-2\pi i\alpha})\ ,
\end{eqnarray}
with  $0 \leqq \alpha, \beta < 1$.
Finally
\begin{equation}
det_{\alpha,\beta}(\bar{\partial}+\nu) = e^{-i\pi \beta \nu}
d
\left[
\begin{array}{c}
                \alpha+\nu  \\
                 \beta    \end{array}  \right] (0|\tau) \ .
\end{equation}
If we evaluate one such component
\begin{eqnarray}
& &\sqrt{det_{\alpha, \beta}
(\bar{\partial}+i\rho_r)det_{-\alpha,-\beta}(\bar{\partial}-i\rho_r)}
\nonumber \\
&=&
e^{\frac{i\pi}{2}(2\alpha\beta-\alpha-\beta)}q^{\frac{(\beta-\frac{1}{2})^2-\frac{1}{12}}{2}}
\prod_{n=1}^{\infty}(1-q^{n-\beta}e^{2\pi
i\alpha}e^{-2\pi\rho_r})(1-q^{n+\beta-1}e^{-2\pi
i\alpha}e^{2\pi\rho_r}) \ , \nonumber
\end{eqnarray}
for $\beta\neq 0$ while for $\beta=0$
\begin{eqnarray}
& &\sqrt{det_{\alpha, 0}
(\bar{\partial}+i\rho_r)det_{-\alpha,0}(\bar{\partial}-i\rho_r)}\nonumber
\\
&=& e^{-\frac{i\pi}{4}\alpha}q^{-\frac{1}{24}}
\prod_{n=1}^{\infty}(1-q^{n}e^{2\pi
i\alpha}e^{-2\pi\rho_r})(1-q^{n}e^{-2\pi
i\alpha}e^{2\pi\rho_r})\nonumber \\
& & \times \sqrt{(1-e^{2\pi
i\alpha}e^{-2\pi\rho_r})(1-e^{-2\pi\alpha}e^{2\pi\rho_r})}\ .
\end{eqnarray}
Note that the zero mode contribution is given by
\begin{equation}
\sqrt{(1-e^{2\pi i\alpha}e^{-2\pi\rho_r})(1-e^{-2\pi
i\alpha}e^{2\pi\rho_r})}=\pm 2i sinh \pi (\rho_r-i\alpha)\ ,
\end{equation}
where $\pm$ sign can be chosen so that it reduces to $|2sin \pi
\alpha|$ with $\rho_r=0$. Given this machinery, one can evaluate the
twisted index of the orbifold
\begin{eqnarray}
& &I(g_1,0;\tau)= \int_{M_{g_1}} dx^i_0 d\psi^i_{-0}
\hat{A}_{M_{g_1}}
\frac{1}{\prod_{\mu=1}^{d}\prod_{n=1}^{\infty}(1-q^{n}e^{-2\pi
\omega_{\mu}})(1-q^{n}e^{2\pi \omega_{\mu}})} \nonumber \\
& &
\frac{\prod_{r=1}^{\frac{d^E}{2}}\prod_{n=1}^{\infty}(1+q^{n-\frac{1}{2}}e^{2\pi
i\alpha_r}e^{-2\pi\rho_r})(1+q^{n-\frac{1}{2}}e^{-2\pi
i\alpha_r}e^{2\pi\rho_r})}
{\prod_{p=1}^{\frac{d_n}{2}}\prod_{n=1}^{\infty}(1-q^{n}e^{2\pi
i\alpha_p}e^{-2\pi\omega_p})(1-q^{n}e^{-2\pi
i\alpha_p}e^{2\pi\omega_p})} \otimes \frac{1}{\pm 2i sinh
\pi(\omega_p-i\alpha_p)}  \nonumber \\
&=& \int_{M_{g_1}} dx^i_0 d\psi^i_{-0} \frac{ det
(1+q^{n-\frac{1}{2}}exp \frac{iF'}{2\pi})}{det (1-q^{n} exp
\frac{iR}{2\pi})det (1-q^{n} exp \frac{iR'}{2\pi})}
\prod_{r=1}^{\frac{d_n}{2}}\frac{1}{\pm 2i sinh
\pi(\omega_p-i\alpha_p)}\ ,
\end{eqnarray}
where $\frac{iR}{2\pi}, \frac{iR'}{2\pi}, \frac{iF'}{2\pi}$ have the
matrices with skew eigenvalues are $\omega_{\mu}, \omega_p- i
\alpha_p, \rho_{r}- i \alpha_r $ , respectively. For $V^{(0)}$
directions, $\alpha_r=0$ is understood. In particular we are
interested in the basic representation and we will read off the
term
of the lowest order in $q$, which is $q^{\frac{1}{2}}$ in this
case,
\begin{equation}
\int_{M_{g_1}} dx^i_0 d\psi^i_{-0} \hat{A}(M_{g_1}) exp
\frac{iF'}{2\pi} \prod _{i=1}^{\frac{d_n}{2}}\frac{1}{\pm 2i sinh
\pi(\omega_p-i\alpha_p)}= {\rm Tr} (-1)^F g_1  \ , \label{AB}
\end{equation}
which is nothing but the Atiyah-Bott fixed point formula. In writing
down the action we discarded the phase factor depending on $\alpha$
and the zero point energy contribution to the fractional factor of
$q$ to match the usual index result. Note that $sinh$ factor gives
the $g_1$ action on the normal component of the curvature.

Now for general twisted sectors
\begin{eqnarray}
& &I(g_1,g_2;\tau)= \int_{M_{g_1g_2}} \hat{A}(M_{g_1g_2})
\prod_{n=1}^{\infty}\frac{det(1+q^{n-\frac{1}{2}}exp
\frac{iF}{2\pi})}{
det (1-q^{n}exp \frac{iR}{2\pi})}  \nonumber \\
& & \otimes
\frac{\prod_{r=1}^{\frac{d_n^E}{2}}\prod_{n=1}^{\infty}(1+q^{n-\beta_r-\frac{1}{2}}e^{2\pi
i\alpha_r}e^{-2\pi\rho_r})(1+q^{n+\beta_r-\frac{1}{2}}e^{-2\pi
i\alpha_r}e^{2\pi\rho_r})}
{\prod_{p=1}^{\frac{d_n}{2}}\prod_{n=1}^{\infty}(1-q^{n-\beta_p}e^{2\pi
i\alpha_p}e^{-2\pi\omega_p})(1-q^{n+\beta_p-1}e^{-2\pi
i\alpha_p}e^{2\pi\omega_p})}   \ .\label{eqtwist}
\end{eqnarray}
Here $F$ is the field strength along $V^{(0)}$ while $d_n^E$ denotes
the dimension of the vector bundle excluding $V^{(0)}$.
The peculiarity of this expression is that there will be contributions
only from the states invariant under the orbifold action. One can
easily check that the projection operators
$\frac{1+g_1+g_1^2+\cdots
g_1^{m-1}}{m}$ for an element $g_1$ of order $m$ acting on the
twisted sector by $g_2$ gives the vanishing contribution to the
twisted index for any state not invariant under the orbifold action.
Thus one of the nontrivial contribution comes from
\begin{equation}
\int_{M_{g_1g_2}}\hat{A}(M_{g_1g_2})q^{\frac{1}{2}}exp
\frac{iF}{2\pi}  \ , \label{twist}
\end{equation}
while more complicated combinations of the tensor representations of
the normal bundle of the tangent bundle and $V^{(r)}$ are possible.
The above term Eq. (\ref{twist}) comes from the twisted states
localized along $M_{g_1g_2}$ and singlet under the rotation group of
the normal directions to $M_{g_1g_2}$. If we consider more general
states transforming nontrivially under the transverse rotation
group, they have more complicated expression, which can be read off
from Eq. (\ref{eqtwist}).

\subsection{Rules for Twisted Sectors}

Since we could in principle have the nontrivial contributions
from the twisted states, we have to decide if we have to worry about
contributions from the lowest twisted states when we are trying to
read off the index for the field theory limit.
We already know that for Euler characteristic we have to
include the contribution from the twisted sectors. We also expect
this is the case for the signature of a four manifold which counts
the difference in the number of self-dual and the anti-self-dual
2-forms  since two forms could arise in the twisted sectors.

To
facilitate the discussion, let us consider a specific
compactification of string theory and figure out the plausible rules
for each index problem. Consider Type IIB compactified on K3. This
is a chiral N=(2,0) theory and the matter contents are completely fixed
by anomaly constraints. Thus for whole moduli space of K3, the
spectrum is the same and this applies for K3 orbifold such as
$T^4/Z_2, T^4/Z_3$. The matter content consists of 1 supergravity
multiplets and 21 tensor multiplets. Supergravity multiplet consists
of graviton and 5 self-dual two-index tensors and gravitinos. One
tensor multiplet consists of 5 scalars and one anti-self-dual tensor
and chiral fermions. The tensors arise from two massless states
$B_{\mu\nu}^i$ with $i=1,2$ of Type IIB string theory, and compactify
self-dual 4-form $B_{\mu\nu\lambda\rho}^+$ on two forms on K3. On K3
we have 3 self-dual 2-forms and 19 anti-self-dual 2-forms  so that
we get 3 self-dual tensors and  19 anti-self-dual tensors out of
$B_{\mu\nu\lambda\rho}^+$. If we realize K3 as $T^4/Z_2$ orbifold
where $Z_2$ acts as $z_1\rightarrow-z_1, \,\, z_2\rightarrow-z_2$,
for a complex coordinate $z_i$ of the torus, we have 16 fixed points
and, upon the blowup, each of the fixed point is replaced by
Eguchi-Hanson space, which supplies one anti-self-dual 2-form. Hence
on each of the twisted sector we obtain one tensor multiplet. Now
from this consideration, it's obvious that we have to include the
twisted sector contribution for the computation of the signature of
K3. This is given by difference of the number of  self-dual tensors
and  anti-self-dual tensors.

Now consider a Dirac spinor of positive chirality spinor $\eta \in
S^+_{K3}$ on K3. If we have such spinors, one obtains the 6-d
gravitino from 10-d gravitino
\begin{equation}
\Psi^M \rightarrow \psi_{\mu}\otimes \eta
\end{equation}
Thus if we have such spinors on K3, one can obtain gravitinos on
6-dimensions as many as those. One can ask if such spinors could
arise from the twisted sectors. But we already saw that on the
twisted sectors, only tensor multiplets arise so that we do not have
gravitinos from the twisted sectors. Had we obtain the gravitinos
from the twisted sectors, we will also have multiple gravitons by
supersymmetry, which is quite weird! Thus spinors of K3 can come
only from untwisted sectors from the $Z_2$ invariant projection of
spinors on $T^4$. These spinors are coming from Ramond sectors of
the right moving sectors\footnote{Spinors could arise from the left
moving sectors for Type II string theory but the arguments goes
parallel to the right-moving modes.} and we can see why the spinors
of K3 could not arise from the twisted sectors. In the untwisted
sector $8_s$ of 10-d postive chiral spinor can be represented as
Ramond vacuum
\begin{equation}
|\pm, \pm,\pm,\pm>\ ,
\end{equation}
in the light cone gauge with total number of  $+$ sign being even.
Here four $\pm$'s denote the eigenvalue of $exp (i\pi J_i)$ where $J_i$
are 4 Cartans of $SO(8)$. Under $SO(4)_1\times
SO(4)_2=SU(2)_{L1}\times SU(2)_{R1}\times SU(2)_{L2}\times
SU(2)_{R2}$ with $SO(4)_2$ action on the transverse K3 directions
\begin{equation}
8^s\rightarrow  (2,1,2,1)\oplus (1,2,1,2) \ ,\label{decomp}
\end{equation}
and under the $Z_2$ orbifold action only $(2,1,2,1)$ is
invariant.\footnote{$Z_2$ action on Ramond vacuum can be written as
$exp(\pi(J_3-J_4))$ where $J_3, J_4$ are the Cartan elements on
$SO(4)_2$.} Combined with $8_v$ of the right moving modes
transforming as a vector under $SO(8)$,
\begin{equation}
8_v\rightarrow (2,2,1,1)\oplus (1,1,2,2)\ ,
\end{equation}
where $(2,2,1,1)$ is invariant under the $Z_2$ action. The
combination
\begin{equation}
(2,1,2,1)\otimes(2,2,1,1) \oplus (1,2,1,2)\otimes (1,1,2,2)
\end{equation}
 give rise to the gravitinos of 6-d after $Z_2$ projection (plus fermions of spin $\frac{1}{2}$).
 Now
consider the twisted sectors. The fermions arise from Ramond vacuum
of right-moving sectors possibly combining with the NS sector states.
Important thing is that Ramond vacuum of the twisted sectors is
given by
\begin{equation}
|\pm,\pm>
\end{equation}
with the total number of $+$ signs  even. Note that this state transforms
nontrivially only under $SO(4)_1$ and is a singlet under $SO(4)_2$.
This could not be a spinor on K3, which transforms nontrivially
under $SO(4)_2$. This argument goes through exactly for the other
$K3$ orbifolds.

For the situation we are interested in, we have to deal with a
spinor in $S^+\otimes E$ with a suitable vector bundle $E$. The
associated Dirac index is naturally defined in the heterotic theory.
Since the orbifold action acting on the right-movers cannot generate
$S^+$ one cannot obtain the desired spinors from the twisted
sectors.

This argument goes through Calabi-Yau 3-fold case as well since
again the Ramond vacuum of the twisted sector is of the form
\begin{equation}
|\pm>  \label{rvacuum}
\end{equation}
and this is a singlet under $SO(6)$ of the transverse direction of
Calabi-Yau 3-fold while spinors on Calabi-Yau 3-fold  transform
nontrivially under $SO(6)$.  Hence in the spacetime supersymmetric
theory, we do not have to consider the contribution from the twisted
sectors for the Dirac index.  This is also true for
nonsupersymmetric orbifold, since these states in the right-moving
modes are the excitation with the tensor representations of the
transverse rotation group around the Ramond vacuum of the type Eq.
(\ref{rvacuum}). This could not match the spinor representation of
the spinors.

Now let's apply this formalism and carry out the Dirac index
computation  for the simplest orbifold $T^4/Z_2$. One can use the
formula Eq. (\ref{AB}). From the projection
\begin{equation}
\frac{1}{2} {\rm Tr}\left( (-1)^F (1+\alpha)\right)
\end{equation}
with $\alpha$ being $Z_2$ action on $T^4$ we find that the first
term $\frac{1}{2}{\rm Tr }(-1)^F$ is vanishing since this counts the Dirac
index on $T^4$. From the $\frac{1}{2} {\rm Tr} (-1)^F g$ one obtains
\begin{equation}
\frac{1}{2} \cdot16 \cdot \frac{1}{(2 sin \frac{\pi}{2})^2}=2 \ ,
\end{equation}
where $16$ comes from the total number of the fixed points and $2
sin \frac{\pi}{2}$ factor comes from the $Z_2$ action on the normal
bundle over the fixed point. Indeed we obtain the right Dirac index
on K3. Similar computation can be done for the other toroidal
orbifold of $K3$ and gives rise to the same result.

\subsection{Orbifold Index  and
the Equivariant Generalization}

Let us come back to our problems of counting BPS bound states of
charge centers and see how the Bose/Fermi orbifolding is reflected
on the equivariant index. Consider the simplest case where we
started with a pair of identical particles and other distinguishable
ones. The orbifolding group would be then $S_2=Z_2$. Here let us
assume that $-\Omega^+=\pm 1$, and then the expected answer, from
either MPS formula or the derivation in the previous section,
is\begin{eqnarray}\label{s2}
\mp(-1)^{\langle2\g_1,\g_2\rangle}\Omega(2\gamma_1+\gamma_2)=
\frac{1}{2!}I(\gamma_1,\gamma_1,\gamma_2)
\pm I(2\gamma_1,
\gamma_2)\frac{1}{2^2} \ .
\end{eqnarray}
On the other hand, we may also read off the index from the
$U(1)$ character-valued index, with the projection operator
\begin{equation}
 \frac{1\pm \alpha}{2} \ ,
\end{equation}
with $\alpha^2=1$. A new ingredient here is the sign in front of $\alpha$,
which chooses whether one retains bosonic or fermionic wavefunctions.

When we apply the formula derived earlier, we have to
know how the bundle of interest would transform under $S_n$ actions
in general. Note that we are dealing with rank one bundle given by
\begin{equation}
{\cal F}=-\frac{1}{2}\sum_{A\neq B} \epsilon_{abc} \,\,
\frac{x^{Ac}-x^{Bc}}{|\vec{x}_A-\vec{x}_B|^3} \,\,q_{AB}
\,\,dx^{Aa}\wedge dx^{Bb}.
\end{equation}
One can check that the vector bundle is invariant under the exchange
of any two $\vec{x}_A, \vec{x}_B$ and hence is invariant under the action of $S_n$.
Reading off individual terms from the formula in the last subsection, the
result for $S_2=Z_2$ is
\begin{eqnarray}
\frac{1}{2}\int_M dx^{\mu}_0 d\psi^{\mu}_{-0} \hat{A}(M)
Ch(F)\pm
\frac{1}{2}\int_{M_2}dx^{i}_0 d\psi^{i}_{-0} \hat{A}(M_2)
Ch(F)\frac{1}{2 i sinh \pi \omega_1'}
\end{eqnarray}
where $M_2$ is the manifold fixed by $Z_2$ action. The first term
produces the first term in Eq. (\ref{s2}), so let us take a look at
the second term.
$\omega_1'=\omega_1-i\alpha_1$ with $\omega_1$ is the eigenvalue of the
curvature tensor along the normal bundle and $2\pi \alpha_1$ is the
rotation angle of the $Z_2$ action, which is $\pi$.
 Here
we assume that the internal
degeneracies are taken into account as explained in the main text.
For the 3 body case with 2 identical particles $\omega_1$
does not contribute for the dimensional reason since
$2 i sinh \pi \omega_1'=2 cosh \pi\omega_1$
gives only even powers of $\omega_1$  and we
have the 2-d fixed manifold with 2-d normal bundle. $sinh $ factor
is due to the $Z_2$ action along the normal bundle, which gives
rise to $\frac{1}{2sin\frac{\pi}{2}}=1/2$. Combined with 1/2 factor
in the projection operator this gives the wanted $\pm{1}/{2^2}$
factor at the second term of Eq. (\ref{s2}).

For $S_3$ orbifold and beyond, it is more convenient to
consider the $U(1)$ equivariant index and use the fixed point
theorem where $U(1)$ is generated by the third component of angular momentum operator
$J_3$. For general action of $g$,\footnote{Here we have $y^{2\pi J_3}$ while in the previous section
we use $y^{2J_3}$ in the index definition. This facilitates the comparison with the corresponding formula appearing in
\cite{Manschot:2011xc}.}
\begin{equation}
{\rm Tr} (-1)^F g y^{2\pi J_3}= \int_{M_g} dx^i_0 d\psi^i_{-0}
\hat{A}(M_g, \nu) Ch (F, \nu) \prod
_{i=1}^{\frac{d_n}{2}}\frac{1}{\pm 2i sinh (\nu L+\pi
\omega_p')}\ ,
\end{equation}
where $\omega_p'=\omega_p-i\alpha_p$ with $\omega_p$ is the
eigenvalue of the
curvature tensor along the normal bundle and $2\pi \alpha_p$ is the
rotation angle and $e^{\nu}=y$.
Following \cite{Manschot:2011xc}, we change the expressions into
equivariant characteristic classes accordingly
and $L$ denotes the action of $J_3$ on $T^{(1,0)}(M_n)$. Now consider $S_2=Z_2$ case. This
equivariant index can be evaluated by evaluating the fixed points
under $J_3$. The fixed points are simply north and south poles of
the sphere, $L$ has the eigenvalue $\pm 1$ and $x_r=0$ assuming we are
dealing with isolated points. Thus,
\begin{equation}
\frac{1}{ i 2sinh (\nu L+\pi \omega_1')}\rightarrow
\frac{1}{2 i sinh (\nu-\frac{\pi i}{2})}=\frac{1}{
2cosh\nu}=\frac{y-y^{-1}}{y^2-y^{-2}}\ .
\end{equation}
Note that the expression is invariant under $\nu\rightarrow
-\nu$
so that we don't have to worry about the sign of $L=\pm 1$.
This persists for all $S_n$ case as well.

 Now consider $S_3$ case.
$S_3$ projection has the form
\begin{equation}
\frac{1}{3!}(1+g+g^2)(1\pm \alpha)\ ,
\end{equation}
where $g=(123)$ meaning $x_1\rightarrow x_2, x_2\rightarrow
x_3,x_3\rightarrow x_1$. Again, we inserted a sign $\pm$ in front
of the order $2$ permutation $\alpha$ to pick out either
bosonic or fermionic wavefunctions. Then $g^2=(132)$ and
$\alpha=(12)$, meaning
$x_1\rightarrow x_2, x_2\rightarrow x_1$. One can check that
$g\alpha=(13), g^2\alpha=(23)$. For the bulk contribution
without $g$ or $\alpha$, the projection operator simply
gives $1/3!$. Now consider
\begin{equation}
\pm \frac{1}{3!}(\alpha+g\alpha+g^2\alpha)\ .
\end{equation}
Each of the three elements give rise to $Z_2$ fixed manifold which
we already worked out. We have the contribution
\begin{equation}
\pm \frac{1}{3!}\cdot 3\cdot
\frac{y-y^{-1}}{y^2-y^{-2}}=\pm
\frac{1}{2}\frac{y-y^{-1}}{y^2-y^{-2}}\ .
\end{equation}
Now turn to $\frac{1}{3!}(g+g^2)$. These two elements are order 3. For each
element we evaluate the equivariant index again by going to a fixed
submanifold. The $\hat{A}$ genus combined with Chern characters give rise to
the usual angular momentum factor. The relevant factor for the
orbifolding effect is
\begin{equation}
\frac{1}{2sinh (\nu+\frac{\pi i}{3})}\cdot\frac{1}{2sinh
(\nu-\frac{\pi i}{3})}=\frac{y-y^{-1}}{y^3-y^{-3}} \ .
\end{equation}
Thus, for the refined case
\begin{eqnarray}
&
&-(-1)^{3\langle\g_1,\g_2\rangle}\Omega_{ref}(3\gamma_1+\gamma_2)=\frac{1}{3!}\CI_{ref}(\gamma_1,\gamma_1,\gamma_1,\gamma_2)
\nonumber \\
& &\hskip 2cm\pm \CI_{ref}(2\gamma_1, \gamma_1,
\gamma_2)\frac{1}{2}\frac{y-y^{-1}}{y^2-y^{-2}}
+\CI_{ref}(3\gamma_1,\gamma_2)\frac{1}{3}\frac{y-y^{-1}}{y^3-y^{-3}}\ ,
\end{eqnarray}
including the factors of the internal degeneracies. Here we are
considering the moduli space of three particle configurations
invariant under $S_3$. Normal directions mean the deformations away
from the fixed configurations in $Z_3$ invariant way. The action is
rotation by $\pm 2\pi/3$. Taking $y=1$ limit, we find
\begin{eqnarray}
-(-1)^{3\langle\g_1,\g_2\rangle}\Omega(3\gamma_1+\gamma_2)=\frac{1}{3!}I(\gamma_1,\gamma_1,\gamma_1,\gamma_2)
\pm I(2\gamma_1, \gamma_1,\gamma_2)\frac{1}{2^2}
+I(3\gamma_1,\gamma_2)\frac{1}{3^2} \ ,
\end{eqnarray}
reproducing the ordinary index formula of the previous section.

It is now obvious how to generalize for general permutation group.
Again we are assuming  $-\Omega^+ =\pm 1$.  If we consider the
various manifolds fixed under the elements of order $m$ the
contribution from the normal directions in the index formula gives
\begin{eqnarray}
&& \frac{1}{2 sinh (\nu +\frac{\pi i}{m})2 sinh (\nu -\frac{\pi
i}{m})2 sinh (\nu +\frac{2\pi i}{m})2 sinh (\nu -2\frac{\pi
i}{m})\cdots}\cr &=& \frac{y-y^{-1}}{y^m-y^{-m}}\ ,
\end{eqnarray}
which becomes $1/m$ in the $y\rightarrow 1$ limit. Consider
$m\beta_1+\beta_2$ for general $m$. One has to figure out the
combinatoric factors appearing in the projection of $S_m$ group
elements. Suppose that $m$ can be written as
\begin{equation}
m=n_1m_1+\cdots +n_lm_l.  \label{partition}
\end{equation}
The number of ways of the partition of $m$ elements into $n_i$ of
$m_i$ elements is given by
\begin{equation}
m! \prod_{i=1}^l \frac{1}{n_i! m_i !}
\end{equation}
For example if we partition $4$ into 2 of 2 elements, we have 3
possibilities, i.e.,
\begin{equation}
12|34, \,\,\, 13|24, \,\,\, 14|23.
\end{equation}
For each particular partition of $m$ elements, one has
$\prod_{i=1}^{l} (m_i-1)!$ permutation elements. For example if we
consider the particular partition of 6 into $123|456$, from $123$ we
have two permutation elements $(123),(132)$ and from $456$ we have
$(456), (465)$ so that one can generate 4 permutation elements out
of this particular partition. Hence the total number of the
permutation elements arising from the partition eq.(\ref{partition})
is
\begin{equation}
m! \prod_{i=1}^l \frac{1}{n_i! m_i}.
\end{equation}
Combined with $\frac{1}{m!}$ in the projection operator of $S_m$
group, and the contribution from the normal directions in the index
theorem, we obtain the factor
\begin{equation}
\prod_{i=1}^l \frac{1}{n_i!}\prod_{i=1}^l
(\frac{1}{m_i}\frac{y-y^{-1}}{y^{m_i}-y^{-m_i}})^{n_i}
\end{equation}
Thus  for simple case of $-\Omega^+ =\pm 1$ we have
\begin{eqnarray}
& & -\Omega_{ref}^{-}(m\beta_1+\beta_2)(-1)^{m<\gamma_1, \gamma_2>}
\nonumber \\
 &=& \sum_{m=\sum_{i=1}^l n_i m_i}\prod_{i=1}^l
\frac{1}{n_i!} I_{ref}(m_1\beta_1, m_1\beta_1, \cdots m_l\beta_l,
m_l\beta_l) \prod_{i=1}^l
(\frac{1}{m_i}\frac{y-y^{-1}}{y^{m_i}-y^{-m_i}}\cdot
-\Omega_1)^{n_i} (-\Omega_2)  \nonumber
\end{eqnarray}
where $I_{ref}$ has $n_i$ of $m_i\beta_i$ factors. The sign of
$-\Omega_i$ can be taken account by the judicious choice of the
signs in the projection operator of $S_m$ as was done for $S_2$ and
$S_3$ cases. Thus we obtain the expected answer.

\section{Summary and Comments }

In this note we showed how $n$ generic BPS dyons of Seiberg-Witten theory
interact with one another, and how the relevant low energy dynamics
with $\CN=4$ supersymmetry can be derived in the vicinity of a wall of
marginal stability. The resulting quantum mechanics is specified by
three classes of quantities: kinetic term, potentials, and minimal couplings.
The latter two turn out to be constrained to each other by supersymmetry
and can be derived exactly, and are universal, in that the
general structure is applicable to BPS black holes as well.
The kinetic term may differ, but for counting
non-threshold bound states via index theorem, we only need the asymptotic
form of the kinetic terms, which fixes effectively the entire Lagrangian.
Thanks to the universal form, this Lagrangian can also be used to compute
non-threshold bound states of BPS black holes as well as those of
Seiberg-Witten dyons.

We showed how the usual truncation (in the previous BPS black hole studies)
down to zero locus, $\CM_n$, of potentials is misleading because the massgaps
along the classically massive direction are always the same as the quantum
massgaps along $\CM_n$, due to the latter's finite size. Instead, one must
sacrifice $\CN=4$ supersymmetry, in favor of an index-preserving $\CN=1$
deformation, in order to reduce the problem to a nonlinear sigma model on
$\CM_n$.

This gives a definite prescription, hitherto unknown, on how to
handle the fermionic superpartners, and the final form of
the index is that of a Dirac operator on $\CM_n$ with an Abelian
gauge field $\CF$ determined unambiguously by the minimal couplings among
dyons/black holes. Along with $n-1$
radial, classical massive directions, $2(n-1)$ fermionic partners become
decoupled from the problem, leaving behind a supersymmetric quantum
mechanics on $\CM_n$ with real supersymmetry. (Three bosonic and four
fermionic variables decouple also, playing the role of the
center of mass degrees of freedom.) This shows rigorously why the
Dirac index is the relevant one, as was anticipated by de Boer et.al. \cite{deBoer:2008zn}.

Since typical wall-crossing problem involves only two linearly
independent charge vectors, and thus  bound states of
many identical BPS states, statistics is of major importance.
We address this directly for  the index problem by inserting
the relevant projection operator $\CP_\Gamma$, and expanding
the index to a series involving various fixed submanifolds.
Each such contribution consists of two multiplicative factors:
one is usual Dirac index on the fixed submanifold and the other is
contribution from the normal direction. The latter turns out
to be universal and generates a numerical factor $\sim 1/p^2$
for each $p$ coincident and identical particles, times the
intrinsic degeneracy of the particle in question. This eventually
lead to the rational invariants,
$\bar \Omega(\g)=\sum_{p\vert \g}\Omega(\g/p)/p^2$, as the
effective degeneracy factor, as was also noted by Manschot et.al. \cite{Manschot:2010qz}.
In the end, we have derived the general  wall-crossing formula,
from the viewpoint of spatially loose BPS bound states
by starting from Seiberg-Witten theory, ab initio.\footnote{
The only unresolved issue here is the angular momentum content
of the bound states, which is in particular needed when one
translates the quantum mechanics index to the second helicity
trace of $D=4$ and $N=2$ field theory. For this, we followed
the well-known assignment, which has been tested in many
explicit examples and widely believed to be correct.}

Another unexpected bonus, which also shows why the low
energy quantum mechanics is never really $2(n-1)$ dimensional
but must be addressed with all $3n$ position coordinates,
was a clarification of various index quantities in literature.
The field theory index is most generally computed by the protected spin character,
$$
{\bf Tr}\left((-1)^{2J_3}y^{2 J_3+2I_3}\right) \ .
$$
We showed how this quantity maps to the quantum mechanical
index
$$
{\rm Tr}\left((-1)^{2J_3}y^{2 J_3+2I_3}\right) \ ,
$$
with the latter's $SO(4)=SU(2)_L\times SU(2)_R$ invariance. This
is then further reduced, mathematically, to the usual equivariant index defined
over $\CM_n$ as
$$
{\rm Tr}\left(\langle0\vert(-1)^{2J_3}\vert 0\rangle_{heavy}y^{2 \CJ_3}\right) \ ,
$$
where $\CJ_3=J_3+I_3$ is naturally picked out because in the
reduction process, which is a purely mathematical operation, we
lose part of the $SO(4)=SU(2)_L\times SU(2)_R$ symmetry. At the
end of the reduction process, only $SU(2)_\CJ$ survives and rotate
bosons and fermions equally, unlike the physical rotation symmetry
$SU(2)_L$ which sees that fermions arise from $D=4$ spinors. Also we
saw how the chirality operator $(-1)^{2J_3}$ is related to the
canonical one $(-1)^{F_{\CM_n}}$ as
$$
{\rm Tr}\left(\langle0\vert(-1)^{2J_3}\vert 0\rangle_{heavy}y^{2 \CJ_3}\right)
=(-1)^{\sum_{A<B}\langle\g_A,\g_B\rangle+n-1}\times {\rm Tr}\left((-1)^{F_{\CM_n}}y^{2 \CJ_3}\right) \ ,
$$
giving us by-now familiar sign in wall-crossing formulae.
This formula is valid before statistics imposed.
Imposing Bose/Fermi statistics introduces a further
complication, which are addressed in much detail in section 5.

An important question that remains unaddressed satisfactorily
here,  or in any other existing literature, is whether
and under what circumstances threshold bound state appear. Primary
examples would be states that are bound state of more than two
of the same BPS states. In all of $D=4$ $N=2$ field theory examples
where explicit states were constructed, no such states ever appeared
as far as we know, yet BPS black holes can be easily found to have
an integer multiple of a given charge.
This tells us that, for addressing such problems, details of the
kinetic term in the low energy dynamics are important and perhaps
one should deal with $3n$ dimensional low energy dynamics directly,
instead of reducing the problem to $\CM_n$; $\CM_n$ would be no longer
compact and thus less useful, since classical ground states would
include configuration with arbitrary small separations and those
with arbitrary large separations among some charge centers. We further
expect that the kinetic
terms for dyons and for black holes, respectively, will be
substantially different at small mutual separations, accounting
the difference. For dyons, this poses additional difficulties since
we must face the fact that dyons in field theory are actually
non-Abelian objects. For black holes, horizon effectively acts
as short-distance cut-off and the configuration remain Abelian
at any separation, and the question of kinetic terms at small
separation is better posed. We hope to return to this problem
in near future.

\vskip 1cm
\centerline{\bf\large Acknowledgement}
\vskip 5mm
We would like to thank Sungjay Lee, Jan Manschot and in particular
Boris Pioline for valuable discussions. P.Y. is grateful to LPTHE
in Paris, Nordita in Stockholm for hospitality, and organizers of
the workshop ``String Phenomenology," where part of this
manuscript was written.
This work is supported  by the National Research Foundation of Korea
(NRF) funded by the Ministry of Education, Science and Technology
with grant number 2010-0013526 (H.K., Z.W.,and P.Y.), and 2009-0085995
(JP) via Basic Science Research Program and also (J.P. and P.Y.) via
the Center for Quantum Spacetime (grant number 2005-0049409). J.P. is
also supported by the KOSEF Grant R01-2008-000-20370-0 and
appreciates APCTP for its stimulating environment for research.

\newpage

\appendix

\centerline{\bf\Large Appendix}

\section{$\CN=4$ Superfield Formalism}
According to \cite{Berezovoj:1991ka} and \cite{Ivanov:1990jn}, one can introduce
$\CN=4$ superfield by
\begin{equation}
\Phi_{\alpha\beta}=(D_{\alpha}\bar{D}_{\beta}+D_{\beta}\bar{D}_{\alpha})
V
\end{equation}
with $V$ being the $D=4$ $N=1$ vector supermultiplet.
Following \cite{Berezovoj:1991ka}, we define\footnote{This is a different
convention from N=1 so we may have to
change the normalization of the fields accordingly to conform toN=1
convention used in the text.}
\begin{eqnarray}
D_{\alpha}&=&
\frac{\partial}{\partial\theta^{\alpha}}-\frac{i}{2}\bar{\theta}_{\alpha}\frac{\partial}{\partial
t} \nonumber \\
\bar{D}^{\alpha}&=&
\frac{\partial}{\partial\bar{\theta}^{\alpha}}-\frac{i}{2}\theta_{\alpha}\frac{\partial}{\partial
t}
\end{eqnarray}
so that
\begin{equation}
\{D_{\alpha}, \bar{D}^{\beta}\}=-i\delta_{\alpha}^{\,\,\,
\beta}\frac{\partial}{\partial t}
\end{equation}
and the indices are raised and lowered by
$\epsilon_{\alpha\beta}=-\epsilon^{\alpha\beta}$.
Alternatively, $\Phi_{\alpha\beta}$ is uniquely determined by
the conditions
\begin{eqnarray}
& &\Phi_{\alpha\beta}=\Phi_{\beta\alpha} \ , \nonumber \\
& &\bar{\Phi}_{\alpha\beta}=\Phi^{\alpha\beta}\ , \nonumber \\
& &
D_{\alpha}\Phi_{\beta\gamma}+D_{\beta}\Phi_{\gamma\alpha}+D_{\gamma}\Phi_{\beta\alpha}=0 \ .
\end{eqnarray}

It is convenient to define real superfields $\hat \Phi$ by
\begin{eqnarray}
\hat\Phi_a&\equiv&
\frac{1}{2}\epsilon^{\beta\gamma}(\sigma_a)_{\gamma}^{\,\,\,
\alpha}\Phi_{\alpha\beta} \nonumber \\
&=&x_a+\frac{i}{2}\theta\sigma_a\bar{\chi}-\frac{i}{2}\chi\sigma_a\bar{\theta}
+\frac{1}{4}\theta\sigma_a\bar{\theta}F
+\frac{1}{2}\epsilon_{abc}\dot{X}_b\theta\sigma_c\bar{\theta}
\nonumber \\
& & +\frac{1}{8}\theta\theta
\bar{\theta}\sigma_a\dot{\bar{\chi}}+\frac{1}{8}\bar{\theta\theta
} \theta\sigma_a\dot{\chi}
+\frac{1}{16}\theta\theta\bar{\theta\theta} \ddot{X}_a \ .
\end{eqnarray}
This defines $ \CN=4$ superfield for 3 bosonic and 4 fermionic
degrees of freedom.
The supersymmetric transformation is given by
\begin{eqnarray}
\delta x_a&=&\frac{i}{2}\epsilon
\sigma_a\bar{\chi}-\frac{i}{2}\chi
\sigma_a\bar{\epsilon}  \nonumber \\
\delta \chi&=& -\epsilon\sigma_a\dot{x}_a+\frac{i}{2} F
\nonumber \\
\delta F&=& \epsilon \dot{\bar{\chi}}-\bar{\epsilon}\dot{\chi}.
\label{susyn=4}
\end{eqnarray}
The kinetic term is given by
\begin{equation}
\CL_0=\frac{1}{2}\int d^4 \theta L(\Phi_a).
\end{equation}
With $f\equiv \partial_a\partial_a L(x_b)$ this can be written
componentwise\footnote{A simple version of this with constant $f$ would be the
dimensional reduction of supersymmetric QED \cite{Smilga:1986rb}.}
\begin{eqnarray}
\CL_0&=&
f(X)(\frac{1}{2}\dot{x_a}^2-\frac{i}{4}\dot{\chi}\bar{\chi}+\frac{i}{4}\chi\dot{\bar{\chi}}+\frac{1}{8}F^2)
\nonumber \\
& &-\frac{1}{8}F\chi\sigma_a\bar{\chi}\partial_a
f+\frac{1}{4}\epsilon_{abc}\dot{X}_b(\chi\sigma_c\bar{\chi})\partial_af-\frac{1}{32}\chi\chi\bar{\chi}\bar{\chi}
\partial_a^2f.
\end{eqnarray}
Now the generalization to $n$ superfield case is straightforward.
One introduces $n$ superfield
$\Phi^A_a, \,\, A=1, \cdots n$. The kinetic part is given by
\begin{equation}
\CL_0=\frac{1}{2}\int d^4\theta L (\Phi^A_a)
\end{equation}
This  action is rather complicated in component form.

Note that in this simple $\CN=4$ superspace formulation,
potential terms are not obvious,\footnote{Nevertheless,
Ref.~\cite{Ivanov:2003tm}
offers a superconformal example with potential terms, written
in an $\CN=4$  off-shell form.}  which is
reasonable since the superfields $\hat \Phi$'s came from
the gauge vector multiplet of $D=4$ by dimensional reduction.
Furthermore, as we emphasized in the main text,
$\CN=1$ superspace description is more convenient since
the evaluation of the index actually relies only on $\CN=1$
supersymmetry. For this,
we split the superfields
$\hat\Phi$ to $3n$ bosonic superfields and $n$ fermionic
superfields as
\begin{equation}
\Phi_a^{A}=x^a_A-i\theta\psi^a_A \ , \,
\Lambda^A=i\lambda^A+i\theta b^A ,\,\,\, A=1,\cdots n, \,\,
a=1,2,3.
\end{equation}
where the complex fermions $\chi$ and $\bar\chi$ are decomposed into
3+1 real fermions $\psi^a$'s and $\lambda$'s, and the auxiliary
field $F$ is now redefined as $b=F/2$.
With these superfields, the general form of the kinetic
Lagrangian $\CL_0$ was worked out in Ref.~\cite{Maloney:1999dv}
\begin{eqnarray}
\CL_0&=&\int d\theta
\biggl(\frac{i}{2}g_{AB}^{ab}D\Phi^A_a\dot{\Phi}^B_b-\frac{1}{2}h_{AB}\Lambda^AD\Psi^B-if_{AB}^a\dot{\Phi}_a^A\Lambda^B
+\frac{1}{3!}c_{ABC}^{abc}D\Phi^A_aD\Phi^B_bD\Phi^C_c \nonumber
\\
&&+\frac{1}{2!}n_{ABC}^{ab}D\Phi^A_aD\Phi^B_b\Lambda^C+\frac{1}{2!}m_{ABC}^a\Phi^A_a\Lambda^B\Lambda^C
+\frac{1}{3!}l_{ABC}\Lambda^A\Lambda^B\Lambda^C\biggr)
\end{eqnarray}
with $D=\frac{d}{d\theta}-i\theta\frac{d}{dt}$, where
\begin{eqnarray}
g_{AB}^{ab}&=&
(\delta^a_d\delta^b_e+\epsilon^{fda}\epsilon_{fe}^{\,\,\,\,\,\,
b})\partial_A^d\partial_B^e L \nonumber \\
h_{AB}&=& \delta_{ab} \partial_A^a\partial_B^b L \nonumber \\
f_{AB}^a&=& \epsilon_{bc}^{\,\,\,\, a}\partial_A^b\partial_B^c
L\nonumber \\
C_{ABC}^{abc}&=&
\frac{1}{2}\epsilon^{pqh}\epsilon_{pl}^{\,\,\,\,
a}\epsilon_{qm}^{\,\,\,\,\,\,\,\, b}\epsilon_{hn}^{\,\,\,\,\,\,
c}
\partial_A^a\partial_B^b\partial_C^c L \nonumber \\
n_{ABC}^{ab}&=&\frac{1}{2}(\epsilon^{pqn}\epsilon_{pl}^{\,\,\,\,a}\epsilon_{qm}^{\,\,\,\,\,\,
b}-\epsilon^l_{\,\,an}\delta_{mb}
-\epsilon^m_{\,\,\,\,bn}\delta_{la})\partial_A^a\partial_B^b\partial_C^c
L \nonumber \\
m_{ABC}^a&=&\frac{1}{2}\epsilon^j_{\,\,
mn}\epsilon_{jl}^{\,\,\,\, a}
\partial_A^a\partial_B^b\partial_C^c L \nonumber \\
l_{ABC}&=& \frac{1}{2}\epsilon_{abc}
\partial_A^a\partial_B^b\partial_C^c L.
\end{eqnarray}
Note that, for the asymptotic form $L(x)=\sum_A m_A(\vec
x_A)^2/2$,
all but the first two, $g$  and $h$, vanish identically.

\section{Reduction to Nonlinear Sigma Model on $\CM_n$}

With $n$ centers, one starts with $3(n-1)$
bosonic coordinates and $4(n-1)$ fermionic ones, after the free center
of mass part is removed from the dynamics. It is convenient to work with a
coordinate system where $n-1$ of them equal to independent linear
combinations of ${\cal K}_A$'s. In a slight abuse of
notation we will denote these again by $\CK^A$, now with $A=1,\dots,n-1$;
although there are $n$ $\CK$'s, only $n-1$ of them are linearly
independent.
Thus, we split the relative part of $r^{Aa}$ and $\psi^{Aa}$
as  $Z^M =({\cal K}^A, y^\mu)$ and $\psi^M =(\psi^A, \psi^\mu)$,
with $M=1,\dots, 3(n-1)$ and $\mu =n,\dots, 3(n-1)$. Along the
same spirit, we also denote by $\lambda^A$, $n-1$ linearly independent
combinations
of $\lambda$'s that belong to the relative part of the low
energy dynamics. What do we mean by $\psi^M$? We wish to preserve
at least one supersymmetry, say $Q_4$, and naturally $\psi^M$ is
the superpartner of $Z^M$,
\begin{equation}
\psi^M=\frac{\partial Z^M}{\partial r^{Aa}}\,\psi^{Aa} \ ,\label{pM}
\end{equation}
and the kinetic term of $\psi^M$ includes two factors of
$\partial r^{Aa}/\partial Z^M$.

As argued in section 4, it suffices to consider the dynamics with
flat metric, which after taking out the center of mass part becomes
$$g_{AaBb}=m_{AB}\,\delta_{ab} \ ,$$
where $m_{AB}$ is the $(n-1)\times(n-1)$ reduced mass matrix.
Expressing this in the curved coordinate system, $Z^M$,
$$g_{MN}=\sum m_{AB}
\frac{\partial r^{Aa}}{\partial Z^M}
\frac{\partial r^{Ba}}{\partial Z^N} \ ,$$
we find that partial derivatives of metric coefficients $g_{MN}$
are nontrivial.
In contrast, nothing much happens to $\lambda$'s, other than
one of them being taken out as the center of mass part, so their
metric is the same reduced mass matrix,
$$h_{AB}=m_{AB} \ ,$$
and is constant. Thus, no coordinate-dependent transformations
are needed for $\lambda$'s. The deformed Lagrangian with flat
kinetic term reads in this coordinate,
\begin{eqnarray}
{\cal L}&=& \frac12\, g_{MN}(Z)\dot Z^M\dot Z^N-
\frac{1}{2}\, \xi^2(m^{-1})^{AB} {\cal K}_A(Z) {\cal K}_B(Z)-{\cal W}(Z)_M\dot Z^M\cr\cr
&+&\frac{i}{2}\, g_{MN}(Z)\psi^M\dot\psi^N-\frac{i}{2}\, \partial_L g_{MN}(Z)\dot Z^N\psi^L\psi^M
+\frac{i}{2}\,m_{AB}\lambda^A\dot\lambda^B
\cr\cr
&+&i\xi \partial_B {\cal
K}_A\psi^B\lambda^A  +i\partial_M {\cal W}_N
(Z)\psi^M\psi^N
\label{ncenter}
\end{eqnarray}
where the crucial middle term in the second line follows
from Eq.~(\ref{pM}) and the anticommuting nature of fermions.
We also used $\partial_\mu \CK_A=0$.

Since we anticipate
that $\CK$ directions will decouple as $\xi\rightarrow \infty$, we split the
metric as
$$
[g_{MN}]=\left(\begin{array}{cc} H_{AB} & C_{A\mu}\\ C^{T}_{\mu A} & G_{\mu\nu}\end{array} \right)
$$
and the likewise for its inverse
$$
[g^{MN}]=\left(\begin{array}{cc} (H- CG^{-1}C^T)^{-1} & -(H-CG^{-1}C^{T})^{-1}C G^{-1}\\
-G^{-1}C^T(H-CG^{-1}C^T)^{-1} & G^{-1}+G^{-1}C^T(H-CG^{-1}C^T)^{-1}C^TG^{-1}\end{array} \right)
$$
Ignoring $\CW$ and fermion contributions to the conjugate momentum for now for simplicity,
the bosonic part of Hamiltonian will then looks something like
\begin{eqnarray}
\CH&\simeq& \frac12g^{\mu\nu}p_\mu p_\nu+ g^{A\mu}p_Ap_\mu+\frac12 g^{AB}p_Ap_B  +\cdots\cr\cr
&=&\frac12 (G^{-1})^{\mu\nu}p_\mu p_\nu+\frac12 g^{AB}P_AP_B+\cdots
\end{eqnarray}
where
$$P_A\equiv p_A+ (H-CG^{-1}C^T)_{AC}g^{C\mu} p_\mu = p_A - (CG^{-1})_A^{\;\;\mu} p_\mu $$
$P_A$'s have the standard canonical commutator with $\CK$'s, so
it is clear that, together with $\sim \xi^2\CK^2$ terms, they form
very heavy harmonic oscillators of frequency $\sim\xi$,  settle down to its
ground state sector, and decouple from ground state counting. This
leaves behind
\begin{eqnarray}\label{step1}
\CH\simeq \frac12 (G^{-1})^{\mu\nu}p_\mu p_\nu+\cdots
\end{eqnarray}
Denoting the canonical conjugate of $p_\mu$ in this reduced dynamics again
by $y^\mu$,\footnote{Generally $\CK$ will mix in the definition of this new
$y^\mu$ coordinates, to reflect the shift of the conjugate momenta,
but this becomes irrelevant because dynamics forces $\CK=0$. Therefore,
the same old $y$ coordinates can be used here.}
the corresponding Lagrangian would be
\begin{eqnarray}
{\cal L}\simeq \frac12\, G_{\mu\nu}\biggl\vert_{\CK=0}\dot y^\mu\dot y^\nu+\cdots
\end{eqnarray}
This makes clear that we could have done the same more simply by imposing $\CK=0$
at the level of Lagrangian.

Procedure leading up to (\ref{step1}) can be repeated in the presence of $\CW$'s,
which simply shift the conjugate momenta in the Hamiltonian,
and it is clear that only $\CW_\mu$'s will survive. We should ask whether this
is consistent, since after all $d\CW$'s are Dirac quantized magnetic fields, and
removing some part of the gauge connection could make the remainder ill-defined.
However, we have
\begin{eqnarray}
d\CW=
\partial_B\CW_A d\CK^B d\CK^A+(\partial_\mu\CW_A -\partial_A\CW_\mu)\,dy^\mu d\CK^A
 +\partial_\nu\CW_\mu dy^\nu dy^\mu\nonumber
\end{eqnarray}
and the pull-back onto $\CM_n$ is simply
\begin{eqnarray}
\CM_n^*(d\CW)=\partial_\nu\CW_\mu dy^\nu dy^\mu
\end{eqnarray}
The pull-back of a well-defined bundle to a smoothly embedded submanifold is still
a well-defined bundle, so the reduced gauge connection $\CW_\mu(\CK=0)$ is consistent. Thus,
the bosonic part of the action reduces to
\begin{eqnarray}
{\cal L}\simeq \frac12\, G_{\mu\nu}\biggl\vert_{\CK=0}
\dot y^\mu\dot y^\nu-\CW_\mu\biggl\vert_{\CK=0}\dot y^\mu+\cdots
\end{eqnarray}
leaving us with the question of how to reduce fermion sector.

The fermions enter the Hamiltonian in two places. One is as
bilinear connection term added to the conjugate momenta, and
the other is an additive contribution of the form
$$
-i\xi \partial_B {\cal
K}_A\psi^B\lambda^A  -i\partial_A{\cal W}_B\psi^A\psi^B
-i(\partial_A{\cal W}_\mu-\partial_\mu \CW_A)\psi^A\psi^\mu
-i\partial_\mu {\cal W}_\nu\psi^\mu\psi^\nu
$$
with the canonical anticommutator among $\psi^M$'s equal to $g^{MN}$.
To disentangle heavy $\psi^A$ from light $\psi^\mu$, we shift
the light fermions as
$$\tilde \psi^\mu\equiv \psi^\mu+ \psi^A(H-CG^{-1}C^T)_{AC}g^{C\mu}
=\psi^\mu - \psi^A (CG^{-1})_A^{\;\;\mu}\ ,$$
such that
$$\{\psi^A,\tilde\psi^\mu\}=0\ , \qquad\{\tilde\psi^\mu,\tilde\psi^\nu\}=(G^{-1})^{\mu\nu} \ .$$
Let us categorize these fermion bilinears into three difference pieces,
$$-i\partial_\mu {\cal W}_\nu\tilde\psi^\mu\tilde\psi^\nu
-i\CE_{A\mu}\psi^A\tilde\psi^\mu+\left[
-i \xi \partial_B {\cal
K}_A\psi^B\lambda^A+\cdots\right]\ .
$$
Terms in the last bracket involve only $\psi^A$ and $\lambda^A$'s
with eigenvalues $\sim\xi$, so these will decouple from the low
energy spectrum. The potential mixing between heavy and light modes
are in
$$
\CE_{A\mu}=\partial_A\CW_\mu-\partial_\mu\CW_A +
 (CG^{-1})_{A}^{\;\;\nu}(\partial_{\nu}\CW_{\mu}-\partial_\mu\CW_\nu)\ .
$$
For heavy sector, this is of course a minor perturbation and ignorable
as $\xi\rightarrow 0$. For light sector, things look less innocent since
the size of this operator is itself not negligible. However, the heavy
fermion enters this operator linearly, and always will connect excited states
and ground states of heavy fermion sector. This forces the energy
eigenvalue differences $(E_n-E_k)$ in the denominator of the perturbation
series to be of order $\sim\xi$, such that the perturbation is suppressed
by powers of $\sim \CE/\xi$. In the end, again, the net effect is to turn
off the heavy modes $\psi^A$ and $\lambda^A$ completely, leaving behind
$$-i\partial_\mu {\cal W}_\nu\psi^\mu\psi^\nu
$$
only, where we will call this light fermion again as $\psi^\mu$'s. The simplest
way to understand this is to recall that any operator linear in heavy fermions
will vanish when sandwiched between heavy sector vacuum.

Combining the reduction processes of the bosonic and the fermionic sectors,
it is clear that the connection term can be equally reduced to
$$\frac{i}{2}\partial_L g_{MN} \dot
Z^N\psi^L\psi^M\quad\rightarrow\quad
\frac{i}{2}\partial_\delta G_{\alpha\beta} \dot
y^\beta\psi^\delta\psi^\alpha \ .$$
Now we can revert from the Hamiltonian to the Lagrangian,
after putting all the heavy modes to their ground states,
and arrive at the following reduced Lagrangian,
\begin{eqnarray}
&&{\cal L}_{\rm for \;index\; only}^{\CN=1}\cr\cr &=& \frac12 G_{\mu\nu}
 (\dot y^\mu \dot y^\nu + i\psi^\mu\dot\psi^\nu )
 -{\CA}_\mu\dot y^\mu
 +\frac{i}{2}\,\CF_{\mu\nu} \psi^\mu\psi^\nu
-\frac{i}{2}\partial_\delta G_{\alpha\beta} \dot
y^\beta\psi^\delta\psi^\alpha \cr\cr
&=& \frac12 G_{\mu\nu} \dot y^\mu \dot y^\nu
+\frac{i}{2}\psi^\mu G_{\mu\nu}(\dot\psi^\nu
+\Gamma^\mu_{\gamma\delta}\dot y^\gamma\psi^\delta) -{\cal
A}_\mu\dot y^\mu +\frac{i}{2}\CF_{\mu\nu} \psi^\mu\psi^\nu \ ,\label{2n}
\end{eqnarray}
where we introduced the notation, also used in the main text,
$\CF=\CM_n^*(d\CW)$ and its gauge field
$\CA$. We already defined $G$ as the appropriate block of $g$,
but now valued at $\CM_n$. In other words, $G =\CM_n^*(g)$.
Remaining fermions live in the co-tangent bundle of ${\cal M}_{n}$, so
the resulting Lagrangian is $\CN=1$ non-linear sigma model
on $2(n-1)$-dimensional manifold ${\cal M}_n$, coupled to an Abelian
gauge field $\CW$. Supercharge of this dynamics is a Dirac
operator on $\CM_n$ coupled to Abelian gauge field $\CA$,
and therefore the index is, under the canonical choice of
the chirality operator,
$$\int_{\CM_n} Ch(\CF)\hat{A}(\CM_n) \ .$$

\end{document}